

\input lanlmac
\input amssym
\input epsf

\newcount\figno
\figno=0
\def\fig#1#2#3{
\par\begingroup\parindent=0pt\leftskip=1cm\rightskip=1cm\parindent=0pt
\baselineskip=13pt
\global\advance\figno by 1
\midinsert
\epsfxsize=#3
\centerline{\epsfbox{#2}}
\vskip 12pt
{\bf Fig. \the\figno:~~} #1 \par
\endinsert\endgroup\par
}
\def\figlabel#1{\xdef#1{\the\figno}}
\newdimen\tableauside\tableauside=1.0ex
\newdimen\tableaurule\tableaurule=0.4pt
\newdimen\tableaustep
\def\phantomhrule#1{\hbox{\vbox to0pt{\hrule height\tableaurule
width#1\vss}}}
\def\phantomvrule#1{\vbox{\hbox to0pt{\vrule width\tableaurule
height#1\hss}}}
\def\sqr{\vbox{%
  \phantomhrule\tableaustep

\hbox{\phantomvrule\tableaustep\kern\tableaustep\phantomvrule\tableaustep}%
  \hbox{\vbox{\phantomhrule\tableauside}\kern-\tableaurule}}}
\def\squares#1{\hbox{\count0=#1\noindent\loop\sqr
  \advance\count0 by-1 \ifnum\count0>0\repeat}}
\def\tableau#1{\vcenter{\offinterlineskip
  \tableaustep=\tableauside\advance\tableaustep by-\tableaurule
  \kern\normallineskip\hbox
    {\kern\normallineskip\vbox
      {\gettableau#1 0 }%
     \kern\normallineskip\kern\tableaurule}%
  \kern\normallineskip\kern\tableaurule}}
\def\gettableau#1 {\ifnum#1=0\let\next=\null\else
  \squares{#1}\let\next=\gettableau\fi\next}

\tableauside=1.0ex
\tableaurule=0.4pt


\def\th{\theta}

\def\vep{\varepsilon}

\def\Tr{{\rm Tr}}
\def\hf{{1\over 2}}
\def\qu{{1\over 4}}

\def\o{\over}

\def\til#1{\widetilde{#1}}
\def\si{\sigma}

\def\b#1{\overline{#1}}
\def\del{\partial}

\def\lap{\Delta}
\def\bra{\langle}
\def\ket{\rangle}
\def\lf{\left}
\def\ri{\right}
\def\riya{\rightarrow}

\def\la{\lambda}

\def\h#1{\widehat{#1}}

\def\bt{\beta}

\def\al{\alpha}
\def\om{\omega}

\def\tens{\otimes}
\def\Om{\Omega}
\def\dag{\dagger}
\def\rt#1{\sqrt{#1}}

\def\sitarel#1#2{\mathrel{\mathop{\kern0pt #1}\limits_{#2}}}
\def\uerel#1#2{{\buildrel #1 \over #2}}

\def\cob{\delta}

\def\F{{\cal F}}
\def\vn{\vec{n}}

\lref\HeckmanBQ{
  J.~J.~Heckman,
  ``Particle Physics Implications of F-theory,''
  arXiv:1001.0577 [hep-th].
}
\lref\HeckmanPV{
  J.~J.~Heckman and H.~Verlinde,
  ``Evidence for F(uzz) Theory,''
  arXiv:1005.3033 [hep-th].
}
\lref\MaldacenaMW{
  J.~Maldacena and D.~Martelli,
  ``The unwarped, resolved, deformed conifold: fivebranes and the baryonic
  branch of the Klebanov-Strassler theory,''
  JHEP {\bf 1001}, 104 (2010)
  [arXiv:0906.0591 [hep-th]].
}
\lref\NastaseNY{
  H.~Nastase, C.~Papageorgakis and S.~Ramgoolam,
  ``The fuzzy $S^2$ structure of M2-M5 systems in ABJM membrane theories,''
  JHEP {\bf 0905}, 123 (2009)
  [arXiv:0903.3966 [hep-th]].
}
\lref\DiaconescuBR{
  D.~E.~Diaconescu, M.~R.~Douglas and J.~Gomis,
  ``Fractional branes and wrapped branes,''
  JHEP {\bf 9802}, 013 (1998)
  [arXiv:hep-th/9712230].
}
\lref\DiaconescuDT{
  D.~E.~Diaconescu and J.~Gomis,
  ``Fractional branes and boundary states in orbifold theories,''
  JHEP {\bf 0010}, 001 (2000)
  [arXiv:hep-th/9906242].
}
\lref\FrancoZU{
  S.~Franco, A.~Hanany, F.~Saad and A.~M.~Uranga,
  ``Fractional Branes and Dynamical Supersymmetry Breaking,''
  JHEP {\bf 0601}, 011 (2006)
  [arXiv:hep-th/0505040].
}
\lref\ButtiHC{
  A.~Butti,
  ``Deformations of toric singularities and fractional branes,''
  JHEP {\bf 0610}, 080 (2006)
  [arXiv:hep-th/0603253].
}
\lref\BerensteinRI{
  D.~Berenstein and R.~Corrado,
  ``Matrix theory on ALE spaces and wrapped membranes,''
  Nucl.\ Phys.\  B {\bf 529}, 225 (1998)
  [arXiv:hep-th/9803048].
}
\lref\DouglasQW{
  M.~R.~Douglas, B.~Fiol and C.~Romelsberger,
  ``The spectrum of BPS branes on a noncompact Calabi-Yau,''
  JHEP {\bf 0509}, 057 (2005)
  [arXiv:hep-th/0003263].
}
\lref\KitazawaXJ{
  Y.~Kitazawa,
  ``Matrix models in homogeneous spaces,''
  Nucl.\ Phys.\  B {\bf 642}, 210 (2002)
  [arXiv:hep-th/0207115].
}
\lref\DymarskyXT{
  A.~Dymarsky, I.~R.~Klebanov and N.~Seiberg,
  ``On the moduli space of the cascading $SU(M+p)\times SU(p)$ gauge theory,''
  JHEP {\bf 0601}, 155 (2006)
  [arXiv:hep-th/0511254].
}
\lref\HoriCD{
  K.~Hori and A.~Kapustin,
  ``Worldsheet descriptions of wrapped NS five-branes,''
  JHEP {\bf 0211}, 038 (2002)
  [arXiv:hep-th/0203147].
}
\lref\AlexanianQJ{
  G.~Alexanian, A.~P.~Balachandran, G.~Immirzi and B.~Ydri,
  ``Fuzzy CP(2),''
  J.\ Geom.\ Phys.\  {\bf 42}, 28 (2002)
  [arXiv:hep-th/0103023].
}
\lref\CarowWatamuraCT{
  U.~Carow-Watamura, H.~Steinacker and S.~Watamura,
  ``Monopole bundles over fuzzy complex projective spaces,''
  J.\ Geom.\ Phys.\  {\bf 54}, 373 (2005)
  [arXiv:hep-th/0404130].
}
\lref\BraunJP{
  V.~Braun, T.~Brelidze, M.~R.~Douglas and B.~A.~Ovrut,
  ``Eigenvalues and Eigenfunctions of the Scalar Laplace Operator on Calabi-Yau
  Manifolds,''
  JHEP {\bf 0807}, 120 (2008)
  [arXiv:0805.3689 [hep-th]].
}
\lref\Ikeda{
A. Ikeda and Y. Taniguchi,
``Spectra and Eigenforms of the Laplacian on $S^n$ and $P^n(C)$,''
Osaka J. Math. {\bf 15} (1978) 515.
}
\lref\GrosseCI{
  H.~Grosse and A.~Strohmaier,
  ``Noncommutative geometry and the regularization problem of 4D quantum  field
  theory,''
  Lett.\ Math.\ Phys.\  {\bf 48}, 163 (1999)
  [arXiv:hep-th/9902138].
}
\lref\SeibergVS{
  N.~Seiberg and E.~Witten,
  ``String theory and noncommutative geometry,''
  JHEP {\bf 9909}, 032 (1999)
  [arXiv:hep-th/9908142].
}
\lref\BerensteinJH{
  D.~Berenstein and R.~G.~Leigh,
  ``Non-commutative Calabi-Yau manifolds,''
  Phys.\ Lett.\  B {\bf 499}, 207 (2001)
  [arXiv:hep-th/0009209].
}
\lref\AspinwallEV{
  P.~S.~Aspinwall,
  ``Resolution of orbifold singularities in string theory,''
  arXiv:hep-th/9403123.
}
\lref\DouglasDE{
  M.~R.~Douglas, B.~R.~Greene and D.~R.~Morrison,
  ``Orbifold resolution by D-branes,''
  Nucl.\ Phys.\  B {\bf 506}, 84 (1997)
  [arXiv:hep-th/9704151].
}
\lref\DouglasSW{
  M.~R.~Douglas and G.~W.~Moore,
  ``D-branes, Quivers, and ALE Instantons,''
  arXiv:hep-th/9603167.
}
\lref\Beilinson{
A.~A.~Beilinson,
``Coherent sheaves on ${\Bbb P}^n$ and problems of linear algebra,''
Funct.\ Anal.\ Appl.\ {\bf 12} (1978), 214-216. 
}
\lref\Bondal{
A.~I.~Bondal,
``Helices, representations of quivers and Koszul algebras,''
in A.~N.~Rudakov et al. {\it Helices and vector bundles} 75-95,
London Mathematical Society Lecture Note Series 148,
Cambridge University Press, 1990.
}
\lref\ProudfootMZ{
  N.~J.~Proudfoot and A.~Bergman,
  ``Moduli spaces for Bondal quivers,''
  Pacific J.\ Math.\  {\bf 237}, 201 (2008)
  [arXiv:math/0512166].
}
\lref\GrosseJT{
  H.~Grosse, C.~Klimcik and P.~Presnajder,
  ``Topologically nontrivial field configurations in noncommutative geometry,''
  Commun.\ Math.\ Phys.\  {\bf 178}, 507 (1996)
  [arXiv:hep-th/9510083].
}
\lref\DolanTX{
  B.~P.~Dolan, I.~Huet, S.~Murray and D.~O'Connor,
  ``Noncommutative vector bundles over fuzzy CP(N) and their covariant
  derivatives,''
  JHEP {\bf 0707}, 007 (2007)
  [arXiv:hep-th/0611209].
}
\lref\KlebanovHH{
  I.~R.~Klebanov and E.~Witten,
  ``Superconformal field theory on threebranes at a Calabi-Yau  singularity,''
  Nucl.\ Phys.\  B {\bf 536}, 199 (1998)
  [arXiv:hep-th/9807080].
}
\lref\GovindarajanVI{
  S.~Govindarajan and T.~Jayaraman,
  ``D-branes, exceptional sheaves and quivers on Calabi-Yau manifolds: From
  Mukai to McKay,''
  Nucl.\ Phys.\  B {\bf 600}, 457 (2001)
  [arXiv:hep-th/0010196].
}
\lref\TomasielloYM{
  A.~Tomasiello,
  ``D-branes on Calabi-Yau manifolds and helices,''
  JHEP {\bf 0102}, 008 (2001)
  [arXiv:hep-th/0010217].
}
\lref\KrippendorfHJ{
  S.~Krippendorf, M.~J.~Dolan, A.~Maharana and F.~Quevedo,
  ``D-branes at Toric Singularities: Model Building, Yukawa Couplings and
  Flavour Physics,''
  arXiv:1002.1790 [hep-th].
}
\lref\IqbalDS{
  A.~Iqbal, N.~Nekrasov, A.~Okounkov and C.~Vafa,
  ``Quantum foam and topological strings,''
  JHEP {\bf 0804}, 011 (2008)
  [arXiv:hep-th/0312022].
}
\lref\HerzogQW{
  C.~P.~Herzog,
  ``Seiberg duality is an exceptional mutation,''
  JHEP {\bf 0408}, 064 (2004)
  [arXiv:hep-th/0405118].
}
\lref\AspinwallVM{
 P.~S.~Aspinwall and I.~V.~Melnikov,
        ``D-branes on vanishing del Pezzo surfaces,''
        JHEP {\bf 0412}, 042 (2004)
        [arXiv:hep-th/0405134].
}
\lref\CecottiZF{
  S.~Cecotti, M.~C.~N.~Cheng, J.~J.~Heckman and C.~Vafa,
  ``Yukawa Couplings in F-theory and Non-Commutative Geometry,''
  arXiv:0910.0477 [hep-th].
}
\lref\DonagiCA{
  R.~Donagi and M.~Wijnholt,
  ``Model Building with F-Theory,''
  arXiv:0802.2969 [hep-th].
}
\lref\BeasleyDC{
  C.~Beasley, J.~J.~Heckman and C.~Vafa,
  ``GUTs and Exceptional Branes in F-theory - I,''
  JHEP {\bf 0901}, 058 (2009)
  [arXiv:0802.3391 [hep-th]];
  ``GUTs and Exceptional Branes in F-theory - II: Experimental Predictions,''
  JHEP {\bf 0901}, 059 (2009)
  [arXiv:0806.0102 [hep-th]].
}
\lref\BeasleyUZ{
  C.~Beasley, B.~R.~Greene, C.~I.~Lazaroiu and M.~R.~Plesser,
  ``D3-branes on partial resolutions of abelian quotient singularities of
  Calabi-Yau threefolds,''
  Nucl.\ Phys.\  B {\bf 566}, 599 (2000)
  [arXiv:hep-th/9907186].
}
\lref\BeasleyZP{
  C.~E.~Beasley and M.~R.~Plesser,
  ``Toric duality is Seiberg duality,''
  JHEP {\bf 0112}, 001 (2001)
  [arXiv:hep-th/0109053].
}
\lref\FengMI{
  B.~Feng, A.~Hanany and Y.~H.~He,
  ``D-brane gauge theories from toric singularities and toric duality,''
  Nucl.\ Phys.\  B {\bf 595}, 165 (2001)
  [arXiv:hep-th/0003085].
}
\lref\FengXR{
  B.~Feng, A.~Hanany and Y.~H.~He,
  ``Phase structure of D-brane gauge theories and toric duality,''
  JHEP {\bf 0108}, 040 (2001)
  [arXiv:hep-th/0104259].
}
\lref\FengBN{
  B.~Feng, A.~Hanany, Y.~H.~He and A.~M.~Uranga,
  ``Toric duality as Seiberg duality and brane diamonds,''
  JHEP {\bf 0112}, 035 (2001)
  [arXiv:hep-th/0109063].
}
\lref\FrancoRJ{
  S.~Franco, A.~Hanany, K.~D.~Kennaway, D.~Vegh and B.~Wecht,
  ``Brane Dimers and Quiver Gauge Theories,''
  JHEP {\bf 0601}, 096 (2006)
  [arXiv:hep-th/0504110].
}
\lref\FengGW{
  B.~Feng, Y.~H.~He, K.~D.~Kennaway and C.~Vafa,
  ``Dimer models from mirror symmetry and quivering amoebae,''
  Adv.\ Theor.\ Math.\ Phys.\  {\bf 12}, 3 (2008)
  [arXiv:hep-th/0511287].
}
\lref\KarabaliIM{
  D.~Karabali and V.~P.~Nair,
  ``Quantum Hall effect in higher dimensions,''
  Nucl.\ Phys.\  B {\bf 641}, 533 (2002)
  [arXiv:hep-th/0203264].
}
\lref\AzumaQE{
  T.~Azuma, S.~Bal, K.~Nagao and J.~Nishimura,
  ``Dynamical aspects of the fuzzy CP(2) in the large N reduced model with  a
  cubic term,''
  JHEP {\bf 0605}, 061 (2006)
  [arXiv:hep-th/0405277].
}
\lref\JanssenCD{
  B.~Janssen, Y.~Lozano and D.~Rodriguez-Gomez,
  ``Giant gravitons and fuzzy CP(2),''
  Nucl.\ Phys.\  B {\bf 712}, 371 (2005)
  [arXiv:hep-th/0411181].
}
\lref\IqbalYE{
  A.~Iqbal, A.~Neitzke and C.~Vafa,
  ``A mysterious duality,''
  Adv.\ Theor.\ Math.\ Phys.\  {\bf 5}, 769 (2002)
  [arXiv:hep-th/0111068].
}
\lref\FrancoES{
  S.~Franco and A.~M.~Uranga,
  ``Dynamical SUSY breaking at meta-stable minima from D-branes at obstructed
  geometries,''
  JHEP {\bf 0606}, 031 (2006)
  [arXiv:hep-th/0604136].
}
\lref\ImamuraFD{
  Y.~Imamura, K.~Kimura and M.~Yamazaki,
  ``Anomalies and O-plane charges in orientifolded brane tilings,''
  JHEP {\bf 0803}, 058 (2008)
  [arXiv:0801.3528 [hep-th]].
}
\lref\BerensteinFI{
  D.~Berenstein and M.~R.~Douglas,
  ``Seiberg duality for quiver gauge theories,''
  arXiv:hep-th/0207027.
}
\lref\FengMI{
  B.~Feng, A.~Hanany and Y.~H.~He,
  ``D-brane gauge theories from toric singularities and toric duality,''
  Nucl.\ Phys.\  B {\bf 595}, 165 (2001)
  [arXiv:hep-th/0003085].
}
\lref\CachazoSG{
  F.~Cachazo, B.~Fiol, K.~A.~Intriligator, S.~Katz and C.~Vafa,
  ``A geometric unification of dualities,''
  Nucl.\ Phys.\  B {\bf 628}, 3 (2002)
  [arXiv:hep-th/0110028].
}
\lref\HananyNM{
  A.~Hanany, C.~P.~Herzog and D.~Vegh,
  ``Brane tilings and exceptional collections,''
  JHEP {\bf 0607}, 001 (2006)
  [arXiv:hep-th/0602041].
}
\lref\KennawayTQ{
  K.~D.~Kennaway,
  ``Brane Tilings,''
  Int.\ J.\ Mod.\ Phys.\  A {\bf 22}, 2977 (2007)
  [arXiv:0706.1660 [hep-th]].
}
\lref\YamazakiBT{
  M.~Yamazaki,
  ``Brane Tilings and Their Applications,''
  Fortsch.\ Phys.\  {\bf 56}, 555 (2008)
  [arXiv:0803.4474 [hep-th]].
}
\lref\HananyVE{
  A.~Hanany and K.~D.~Kennaway,
  ``Dimer models and toric diagrams,''
  arXiv:hep-th/0503149.
  }
\lref\FrancoRJ{
  S.~Franco, A.~Hanany, K.~D.~Kennaway, D.~Vegh and B.~Wecht,
  ``Brane Dimers and Quiver Gauge Theories,''
  JHEP {\bf 0601}, 096 (2006)
  [arXiv:hep-th/0504110].
}
\lref\VerlindeJR{
  H.~Verlinde and M.~Wijnholt,
  ``Building the Standard Model on a D3-brane,''
  JHEP {\bf 0701}, 106 (2007)
  [arXiv:hep-th/0508089].
}
\lref\GopakumarZD{
  R.~Gopakumar, S.~Minwalla and A.~Strominger,
  ``Noncommutative solitons,''
  JHEP {\bf 0005}, 020 (2000)
  [arXiv:hep-th/0003160].
}
\lref\MarchesanoRZ{
  F.~Marchesano and L.~Martucci,
  ``Non-perturbative effects on seven-brane Yukawa couplings,''
  Phys.\ Rev.\ Lett.\  {\bf 104}, 231601 (2010)
  [arXiv:0910.5496 [hep-th]].
}
\lref\SaemannGF{
  C.~Saemann,
  ``Fuzzy Toric Geometries,''
  JHEP {\bf 0802}, 111 (2008)
  [arXiv:hep-th/0612173].
}

\Title{             
                                              }
{\vbox{
\centerline{D-branes Wrapped on Fuzzy del Pezzo Surfaces}
}}

\vskip .2in

\centerline{Kazuyuki Furuuchi${}^1$ and Kazumi Okuyama${}^2$}
\vskip5mm
\centerline{${}^1$National Center for Theoretical Sciences}
\centerline{National Tsing-Hua University, Hsinchu 30013, Taiwan,
R.O.C.}
\centerline{\tt furuuchi@phys.cts.nthu.edu.tw}
\vskip3mm

\centerline{${}^2$Department of Physics, Shinshu University}
\centerline{Matsumoto 390-8621, Japan}
\centerline{\tt kazumi@azusa.shinshu-u.ac.jp}
\vskip .2in

\vskip 2cm
\noindent

We construct classical solutions
in quiver gauge theories on D0-branes probing
toric del Pezzo singularities in Calabi-Yau manifolds. 
Our solutions
represent D4-branes wrapped around fuzzy del Pezzo
surfaces. We study the fluctuation spectrum
around the fuzzy ${\Bbb C}{\Bbb P}^2$ solution in detail.
We also comment on possible applications of our fuzzy
del Pezzo surfaces
to the fuzzy version of F-theory, dubbed F(uzz) theory.

\Date{August 2010}

\vfill
\vfill
\newsec{Introduction}
It is well known that the worldvolume theory on D-branes
becomes non-commutative when the non-zero $B$-field
is threading the cycle on which the D-branes are wrapped \SeibergVS.
Recently, it was argued that 
this type of non-commutative deformation
of D-brane worldvolume theory may have interesting implication
for the phenomenology of F-theory GUT models 
\refs{\CecottiZF,\MarchesanoRZ,\HeckmanPV}.
In \HeckmanPV, the F-theory GUT model based on the 
7-branes wrapped on a fuzzy 4-cycle 
in a Calabi-Yau manifold
was dubbed ``F(uzz) theory'',
and the fuzzy geometry was analyzed by
quantizing the data of gauged linear sigma model.
In the local approach of
F-theory GUTs 
\refs{\DonagiCA,\BeasleyDC} (see \HeckmanBQ\ for a review)
where
the gravity can in principle be decoupled 
by taking the 4-dimensional Planck mass $M_{pl}$ to infinity,
which amounts to considering a non-compact Calabi-Yau manifold,
the possible configurations of 7-brane are quite restricted. 
It turns out that the 7-brane should wrap on a del Pezzo surface
inside the Calabi-Yau manifold.
Therefore, it will be interesting to analyze the D-branes wrapped 
around fuzzy del Pezzo surfaces.

In this paper, we construct 
fuzzy del Pezzo surfaces 
directly as classical solutions of quiver gauge theories
which appear as worldline theories of D0-branes probing 
the del Pezzo surfaces.\foot{%
In our fuzzy del Pezzo surfaces,
the non-commutativities are between
holomorphic- and anti-holomorphic coordinates.
Non-commutative Calabi-Yau manifolds with
non-commutativities between holomorphic coordinates
have been studied previously by some authors (see e.g. \BerensteinJH).}
Our solutions represent a collection of D0-branes puffed up into
D4-branes which 
wrap around a fuzzy del Pezzo surface.
In our construction,
some of the bi-fundamental fields,
which correspond to arrows in the quiver diagram,
represent the non-commutative coordinates of
the del Pezzo surface. 
These non-commutative coordinates
are realized as harmonic oscillators 
whose Fock space corresponds 
to a quantized version of
%
the toric diagram of the del Pezzo surface.
The $k$-th del Pezzo surface $dP_k$, the blow-up of ${\Bbb C}{\Bbb P}^2$
at $k$ generic points, is toric only up to $k=3$,
hence we will
consider classical solutions describing 
D4-branes wrapped on fuzzy
$dP_k$ with $k=0,1,2,3$.
These solutions are constructed
by taking the ranks of the gauge group
at the nodes of the quiver different from each other,
and this amounts to considering fractional branes 
\refs{\DiaconescuBR,\DiaconescuDT,\DouglasQW,\FrancoZU,\ButtiHC}
probing the 
del Pezzo singularity.

Our study is complimentary  
to the previous studies 
on the moduli space of quiver gauge theories 
realized 
on a single D-brane
probing a toric singularity in a Calabi-Yau manifolds.
In this case,
the moduli space of the quiver gauge theory
coincides with the Calabi-Yau manifold itself 
\refs{\DouglasSW,\DouglasDE,\BeasleyUZ}.
On the other hand, our D4-brane solutions wrap on
a (fuzzy) four-cycle in the Calabi-Yau manifold,
and typically
they do not have moduli corresponding to 
moving in the directions
in the Calabi-Yau manifold.\foot{%
Some of our solutions have some moduli left, 
as discussed in section 5.}

Branes wrapped on
fuzzy ${\Bbb C}{\Bbb P}^2$ have been studied in the context
of IIB matrix model \KitazawaXJ\ 
or as a giant graviton \JanssenCD.
Our construction of fuzzy ${\Bbb C}{\Bbb P}^2$,
or more generally fuzzy del Pezzo surfaces,
is different from those studied in these references, 
in particular we start with a quiver gauge theory which does
not contain Chern-Simons terms which could 
support the fuzzy 4-cycle via
the Myers effect.
Instead, our solutions use the bi-fundamental fields and
can be regarded as a generalization of a 
similar construction of fuzzy spheres in
\refs{\BerensteinRI,\NastaseNY,\MaldacenaMW}. 
Our construction also has a relation to 
the Beilinson's construction
of stable vector bundles on ${\Bbb C}{\Bbb P}^2$ \Beilinson.
The Beilinson quiver is obtained by deleting some of the arrows in
the McKay quiver.
In our solutions, some of the bi-fundamental fields
are set to zero, which corresponds to
deleting the arrows in the McKay quiver.

If we naively generalize our construction of fuzzy
D4-branes and try to construct 
D7-branes wrapped on a del Pezzo surface
by starting from the worldvolume theory of D3-branes
instead of D0-branes,
we encounter the problem of gauge anomaly in
the 4-dimensional theory on the D3-branes,
since 
the ranks of the gauge group 
at different nodes are different. 
We will comment on how to circumvent this problem.
After that, we will consider a possible application
of our construction of D7-branes
wrapped on a fuzzy del Pezzo surface to F(uzz) theory.

This paper is organized as follows.
In section 2 we review the construction
of D2-branes wrapped on fuzzy ${\Bbb C}{\Bbb P}^1$
as classical solutions of $A_1$ quiver gauge theories,
as a warm-up for our study of 
D4-branes wrapped on fuzzy del Pezzo surfaces.
In section 3 we construct classical solutions of
quiver gauge theories on 
D0-branes probing ${\Bbb C}^3/{\Bbb Z}_3$ orbifold.
Our solutions represent D4-branes wrapped on fuzzy ${\Bbb C}{\Bbb P}^2$.
In section 4 we study the KK spectrum on fuzzy ${\Bbb C}{\Bbb P}^2$
and show that the higher KK modes are truncated by the effect of
non-commutativity on fuzzy ${\Bbb C}{\Bbb P}^2$.
In section 5, by generalizing the construction of fuzzy ${\Bbb C}{\Bbb P}^2$
we construct classical solutions 
corresponding to D4-branes wrapped on
fuzzy $dP_k$ for $k=1,2,3$.
In section 6 we briefly mention possible applications of
our fuzzy del Pezzo surfaces to F(uzz) theory, and we
comment on the issue of gauge anomaly in the quiver gauge theories in 4 dimensions.
Finally, we conclude with discussions in section 7.
In addition, two appendices are included in this paper:
Appendix A is a review of the computation
of D-brane charges for the ${\Bbb C}^3/{\Bbb Z}_3$ quiver theory.
In appendix B we discuss intersections of curves
on fuzzy ${\Bbb C}{\Bbb P}^2$.

\newsec{Review of D-Branes Wrapped on Fuzzy ${\Bbb C}{\Bbb P}^1$}
In \BerensteinRI,
fractional D-branes wrapped on ${\Bbb C}{\Bbb P}^1$ 
in the $A_1$ ALE space were studied as 
classical solutions of the $A_1$ quiver gauge theory.
It turned out that the
resulting ${\Bbb C}{\Bbb P}^1$
was non-commutative, reflecting the matrix
nature of the underlying gauge theory. 
A similar configuration of D5-brane wrapping fuzzy
${\Bbb C}{\Bbb P}^1$ 
also appears in the Klebanov-Witten
theory \refs{\DymarskyXT,\MaldacenaMW},
which describes the worldvolume theory 
of D3-branes on 
the conifold singularity \KlebanovHH. 
In this section, we review the classical
solution of $A_1$ quiver theory
representing the D-branes wrapped on fuzzy ${\Bbb C}{\Bbb P}^1$
as a warm up for the study of
D4-branes wrapping a fuzzy 4-cycle.

The gauge theory on the D-branes probing
the ${\Bbb C}^2/{\Bbb Z}_2$ 
orbifold singularity 
is summarized by the $A_1$ quiver diagram \DouglasSW.
Namely, the gauge group is $U(n_1)\times U(n_2)$ 
associated to the two nodes
of the $A_1$ quiver diagram and the arrows connecting the nodes 
represent the bi-fundamental chiral multiplets
$A_i, B_i~(i=1,2)$ which 
transform under the gauge group
as $(n_1,\b{n_2})$ and $(\b{n_1},n_2)$, respectively. 
The quiver gauge theory for the pure D0-branes
on the ${\Bbb C}^2/{\Bbb Z}_2$ orbifold
corresponds to the gauge group
of equal rank for the two nodes, while
the fractional branes wrapping the 2-cycle
${\Bbb C}{\Bbb P}^1$ 
is described by the same quiver gauge theory
with unequal rank $n_1\not=n_2$.

Let us consider the (classical) vacuum of this quiver gauge theory.
The D-term condition is
\eqn\DeqAone{\eqalign{
\sum_{i=1,2}(A_i A_i^\dag-B_i^\dag B_i)&=\zeta_1{\bf 1}_{n_1} \cr
\sum_{i=1,2}(B_iB_i^\dag-A_i^\dag A_i)&=\zeta_2{\bf 1}_{n_1}~.
}}
The F-term condition is solved by simply setting
$B_1=B_2=0$. Then the above D-term condition becomes
\eqn\DcondAoneforA{
\sum_{i=1,2}A_i A_i^\dag=\zeta_1{\bf 1}_{n_1},\quad
-\sum_{i=1,2}A_i^\dag A_i=\zeta_2{\bf 1}_{n_2}~.
}
To find a solution that represents a D2-brane wrapped on fuzzy ${\Bbb C}{\Bbb P}^1$, 
it is convenient
to introduce two independent oscillators $a_1,a_2$
obeying the usual commutation relations
\eqn\oscidef{
[a_i,a_j^\dag]=\cob_{ij},\quad
[a_i,a_j]=[a_i^\dag,a_j^\dag]=0\qquad(i=1,2)~,
}
and we take the Chan-Paton vector spaces
${\Bbb C}^{n_1},~{\Bbb C}^{n_2}$
as 
\eqn\CnasFN{
{\Bbb C}^{n_1}={\cal F}^{(2)}_N,\quad
{\Bbb C}^{n_2}={\cal F}^{(2)}_{N+1}~.
}
Here ${\cal F}_N^{(2)}$
denotes the Fock space of two oscillators
with the total occupation number fixed to $N$:
\eqn\foxkVN{
{\cal F}^{(2)}_N=\left\{|m_1,m_2\ket={(a_1^\dag)^{m_1}(a_2^\dag)^{m_2}\over
\sqrt{m_1!m_2!}}|0\ket,~~m_1+m_2=N\right\}~.
}
For our choice of 
Chan-Paton spaces \CnasFN, 
the ranks of two gauge groups $n_1,n_2$ 
are given by
\eqn\nifortwo{
n_1={\rm dim}{\cal F}^{(2)}_N=N+1,\quad
n_2={\rm dim}{\cal F}^{(2)}_{N+1}=N+2~.
}
Then the bi-fundamental matter
$A_i~(i=1,2)$ can be viewed as a linear map from ${\cal F}^{(2)}_{N+1}$
to ${\cal F}^{(2)}_N$:
\eqn\Aiasmatinone{
{\cal F}^{(2)}_{N+1}~\uerel{A_i}{\longrightarrow}~{\cal F}^{(2)}_N~.
}
We can easily find a solution of \DcondAoneforA\
by identifying $A_i$ as the annihilation operator
$a_i$ whose action is restricted to the space ${\cal F}^{(2)}_{N+1}$:
\eqn\Aiaisol{
A_i=ca_i\Big|_{{\cal F}^{(2)}_{N+1}}~,
} 
where $c$ is a complex number.
The relation between the constant $c$ 
and the FI-parameters
$\zeta_r$ $(r=1,2)$ is determined from \DcondAoneforA\foot{When 
the gauge group is $SU(n_1)\times SU(n_2)$
as in the Klebanov-Witten theory, $|c|^2$ is not a parameter but the VEV
of baryonic current $\Tr(A_iA_i^\dag-B_i^\dag B_i)$ \refs{\DymarskyXT,\MaldacenaMW}.}.
Using the relation\foot{Throughout 
this paper we use the relation $[a_i,a_j^\dagger]=\cob_{ij}$ restricted
to a finite dimensional space such as ${\cal F}_N^{(2)}$.
This relation holds exactly on the finite dimensional
space if we read it as $a_i\big|_{{\cal F}_{N+1}^{(2)}} 
a_j^\dag\big|_{{\cal F}_{N}^{(2)}}-a_j^\dag\big|_{{\cal F}_{N-1}^{(2)}} 
a_i\big|_{{\cal F}_{N}^{(2)}}=\cob_{ij}{\bf 1}_{{\cal F}_N^{(2)}}$.
In other words, this relation has the form
$M_iN_j-\til{N}_j\til{M}_i=\cob_{ij}$ where 
$M_i,\til{M}_i,N_j,\til{N}_j$ are finite dimensional matrices.
Note that the usual argument
for the impossibility of realizing the relation $[a_i,a_j^\dag]=\cob_{ij}$
in a finite dimensional space by taking trace on both sides does 
not apply here since $M_i\not=\til{M}_i, N_j\not=\til{N}_j$. 
}
\eqn\twoairel{
\sum_{i=1}^2a_ia_i^\dag=2+\sum_{i=1}^2a_i^\dag a_i~,
}
we find that the coefficient $c$ 
in the solution \Aiaisol\
and the FI-parameters are related by
\eqn\cvszetaone{
\zeta_1=|c|^2(N+2),\quad \zeta_2=-|c|^2(N+1)~.
}
One can check that
the overall $U(1)$ of the gauge group $\prod_r U(n_r)$
is decoupled since
\eqn\sumFItwo{
\sum_{r=1}^2 n_r\zeta_r=(N+1)\zeta_1+(N+2)\zeta_2=0~.
}

To see this solution describes fuzzy 
${\Bbb C}{\Bbb P}^1$,
let us introduce generators of $su(2)$:
\eqn\Jatwo{
J^a=\hf a_i^\dag\si^a_{ij}a_j~,
}
where $\si^a~(a=1,2,3)$ denote the $2\times2$ Pauli matrices.
We can easily show that the above $J^a$ satisfy the
$su(2)$ algebra $[J^a,J^b]=i\varepsilon^{abc}J^c$,
and the Casimir element is given in terms of 
the total number of oscillators $\h{N}=a_1^\dag a_1+a_2^\dag a_2$:
\eqn\sutwoCasimir{
\sum_{a=1}^3J^aJ^a=\qu \h{N}(\h{N}+2)~.
}
Fuzzy ${\Bbb C}{\Bbb P}^1$ is defined by the matrix algebra 
acting on the Fock space ${\cal F}^{(2)}_N$, i.e.
\eqn\defCPoneN{
{\Bbb C}{\Bbb P}^1_N={\rm End}({\cal F}^{(2)}_N)=
{\rm Mat}
\lf(N+1,{\Bbb C}\ri)~,
}
where ${\rm Mat}(d,{\Bbb C})$ denotes the space of $d\times d$ matrices.
Note that 
$J^a$ in \Jatwo\ restricted on ${\cal F}^{(2)}_N$ generates the algebra
of ${\Bbb C}{\Bbb P}^1_N$. More generally,
the space of linear maps ${\rm Hom}({\cal F}_N^{(2)},{\cal F}_{N+m}^{(2)})$
can be thought of as the quantization of the sections of
the line bundle ${\cal O}_{{\Bbb P}^1}(m)$.

To summarize,
the solution in \Aiaisol\ represents a D2-brane wrapped on fuzzy 
${\Bbb C}{\Bbb P}^1$, which can also be viewed as 
a collection of D0-branes puffed up into the D2-brane.
As emphasized in \HeckmanPV, the Fock space
${\cal F}_N^{(2)}$ can be viewed as a set
of ``fuzzy points'' on ${\Bbb C}{\Bbb P}^1_N$.

\newsec{D-branes wrapped on fuzzy ${\Bbb C}{\Bbb P}^2$}
In this section, generalizing the construction 
of fuzzy ${\Bbb C}{\Bbb P}^1$
in the previous section,
we will construct classical solutions of the quiver gauge theory
on the D0-branes probing a
${\Bbb C}^3/{\Bbb Z}_3$ orbifold singularity,
and show that our solutions
represent D4-branes wrapped on fuzzy ${\Bbb C}{\Bbb P}^2$.

\subsec{Quiver Gauge Theory for ${\Bbb C}^3/{\Bbb Z}_3$ Orbifold}
The worldline theory on the D0-branes probing 
${\Bbb C}^3/{\Bbb Z}_3$ singularity is summarized
by the quiver diagram of three nodes with
the gauge group $U(n_1)\times U(n_2)\times U(n_3)$ (see Fig. 1a)
\refs{\DiaconescuDT,\DouglasQW}.
Each pair of nodes are connected by three arrows,
which correspond to the
bi-fundamental chiral multiplets
in 4-dimension dimensionally reduced to 1-dimension.
We denote those chiral multiplets as
$A_i,B_i,C_i~(i=1,2,3)$, and they transform
under the gauge group as 
\eqn\ABCrep{\eqalign{
A_i&\in (n_1,\b{n_2},1)\cr
B_i&\in (1,n_2,\b{n_3})\cr
C_i&\in (\b{n_1},1,n_3)~. 
}}
The orbifold ${\Bbb C}^3/{\Bbb Z}_3$ can be thought of as a
degenerate limit of the non-compact Calabi-Yau space
${\cal O}_{{\Bbb P}^2}(-3)$ where the base ${\Bbb C}{\Bbb P}^2$ 
collapsed to a point. Therefore, we expect that
the fractional D4-branes wrapped on vanishing
${\Bbb C}{\Bbb P}^2$ can be described as a classical solution
of this quiver gauge theory
with an appropriate choice of
$n_{1,2,3}$.
\fig{(a) The McKay quiver for the worldline theory
of D0-branes on ${\Bbb C}^3/{\Bbb Z}_3$ orbifold.
(b) The Beilinson quiver describing
vector bundles on ${\Bbb C}{\Bbb P}^2$ is obtained by setting $C_i=0$
in the diagram (a). 
}{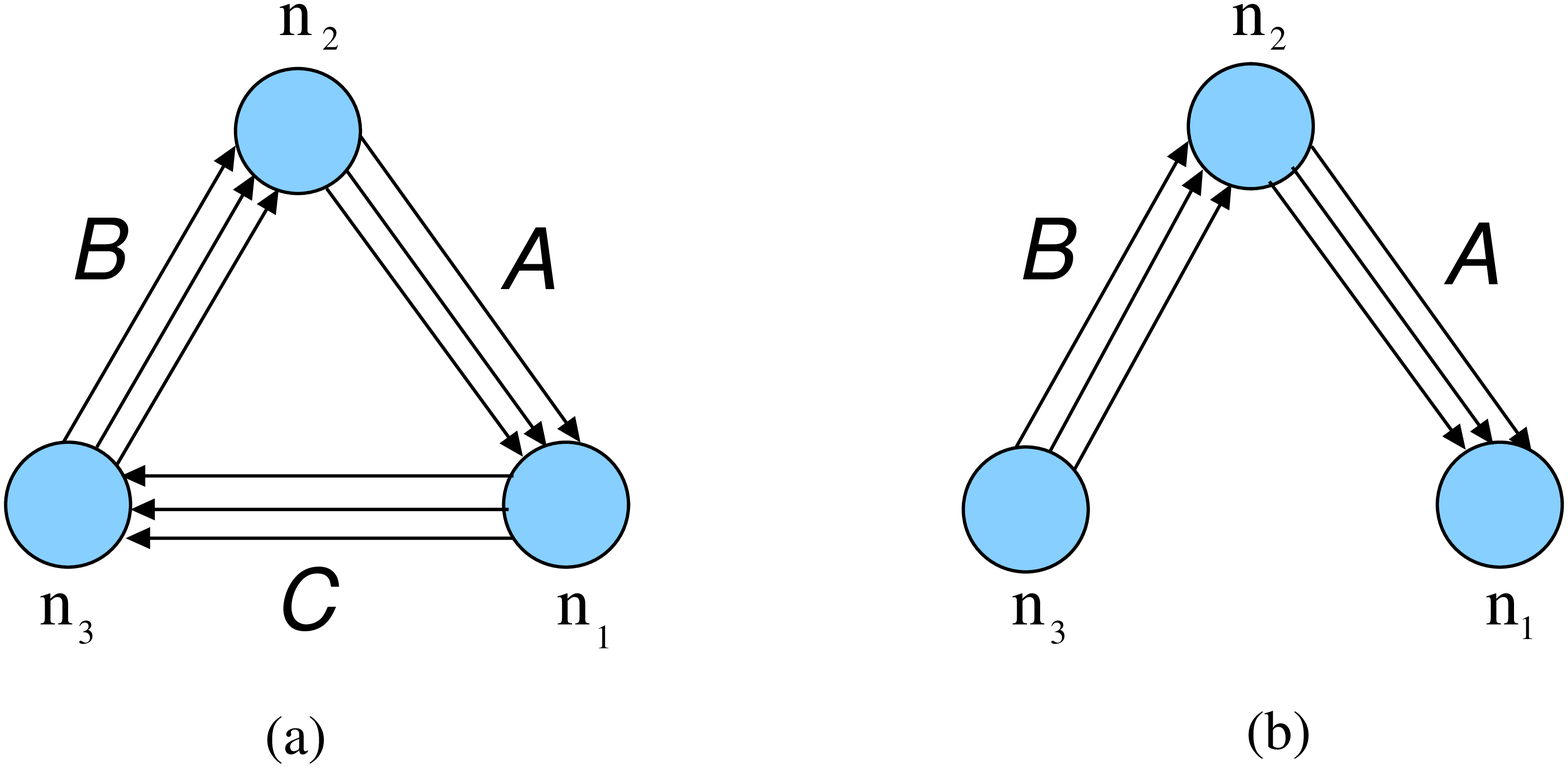}{7cm}

Let us study the classical vacuum of this quiver gauge theory.
The F-term condition coming from the superpotential
$W=\vep^{ijk}\Tr(A_iB_jC_k)$ reads
\eqn\FABcom{
A_iB_j=A_jB_i,\quad
B_iC_j=B_jC_i,\quad
C_iA_j=C_jA_i\quad(i\not=j)~,
}
and the D-term condition is
\eqn\DcondAB{\eqalign{
\sum_{i=1}^3(A_i A_i^\dag-C_i^\dag C_i)&=\zeta_1{\bf 1}_{n_1}\cr
\sum_{i=1}^3(B_iB_i^\dag-A_i^\dag A_i)&=\zeta_2{\bf 1}_{n_2}\cr
\sum_{i=1}^3(C_iC_i^\dag-B_i^\dag B_i)&=\zeta_3{\bf 1}_{n_3}~.
}
}
The last two equations in
the F-term condition \FABcom\ are satisfied by setting $C_i=0$ for all
$i=1,2,3$.
This 
can be represented by deleting the arrows connecting the nodes 1 and 3.
The resulting quiver diagram with no arrows between
the 
nodes 1 and 3 is known as the Beilinson quiver (see Fig. 1b), which was
used to characterize the stable vector bundles on ${\Bbb C}{\Bbb P}^2$ 
\refs{\Beilinson}.
After setting $C_i=0$, the 
F-term and D-term
conditions for $A_i,B_i$ become
\eqn\reducedABeq{\eqalign{
&\qquad A_iB_j=A_jB_i\quad(i\not=j),\cr
\sum_{i=1}^3A_i A_i^\dag=\zeta_1{\bf 1}_{n_1},\quad
&\sum_{i=1}^3(B_iB_i^\dag-A_i^\dag A_i)=\zeta_2{\bf 1}_{n_2},
\quad -\sum_{i=1}^3B_i^\dag B_i=\zeta_3{\bf 1}_{n_3}~.
}}
In the next subsection, we will consider solutions of
this set of equations. 

\subsec{Solutions from the Beilinson Quiver}
We can consider various solutions of \reducedABeq\
by choosing $n_{1,2,3}$ appropriately.
One interesting solution of \reducedABeq\ is found by taking
\eqn\ninN{
{\Bbb C}^{n_1}={\cal F}^{(3)}_{N},\quad
{\Bbb C}^{n_2}={\cal F}^{(3)}_{N+1},\quad
{\Bbb C}^{n_3}={\cal F}^{(3)}_{N+2},
}
where ${\cal F}^{(3)}_{N}$ denotes
the Fock space of three independent oscillators
with total occupation number $N$:
\eqn\VNthree{
{\cal F}_N^{(3)}
=\left\{|m_1,m_2,m_3\ket={(a_1^\dag)^{m_1}(a_2^\dag)^{m_2}(a_3^\dag)^{m_3}\over
\sqrt{m_1!m_2!m_3!}}|0\ket,~~m_1+m_2+m_3=N\right\}~.
}
The dimension of this space is
\eqn\dinFNthree{
{\rm dim}{\cal F}^{(3)}_N=\hf(N+1)(N+2)~.
}
\dinFNthree\ can be counted from 
Fig. 2,
which is a quantized version of 
the toric diagram of ${\Bbb C}{\Bbb P}^2$.
See Figure 3 of \IqbalYE\ for the toric diagram
of commutative ${\Bbb C}{\Bbb P}^2$.
\fig{``Quantized'' toric diagram of fuzzy ${\Bbb C}{\Bbb P}^2$.
${\Bbb C}{\Bbb P}^2$ is described as 
$S^1 \times S^1$ fiber over the base space
represented by the triangle of the toric diagram.
The black points represent the points of the ``quantized'' base space
of fuzzy ${\Bbb C}{\Bbb P}^2$.
The total number of the points in the quantized base space is finite.
}{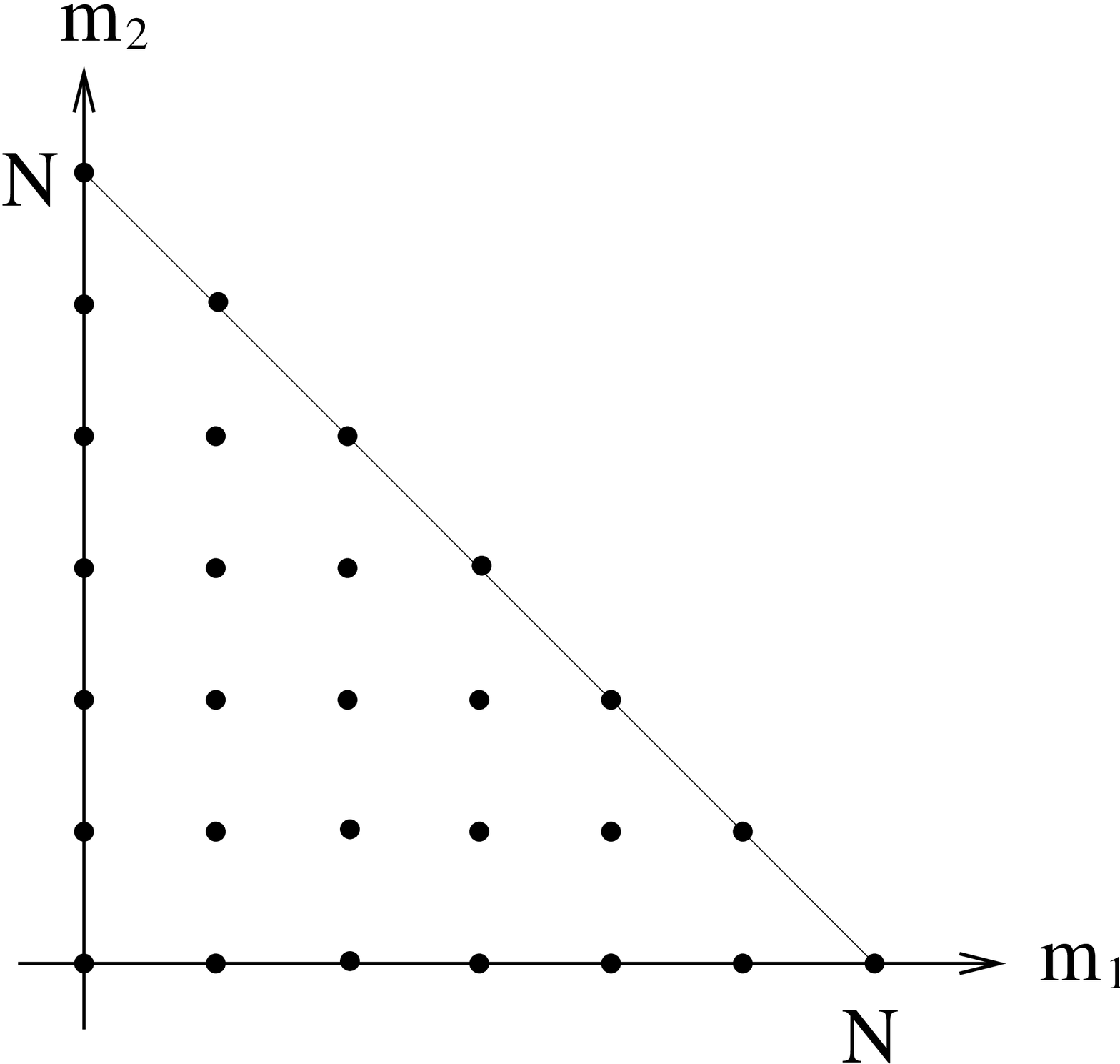}{5cm}
\noindent The dimension of ${\cal F}^{(3)}_N$ 
is given by the total number of the black points in Fig.2.
As emphasized in \HeckmanPV, 
fuzzy ${\Bbb C}{\Bbb P}^2$ 
(or more precisely the base space when one sees
${\Bbb C}{\Bbb P}^2$ as a torus fibered over the base space,
as we will explain shortly)
is made of finite number of points,
which has many interesting consequences.
Let us recall the description of commutative ${\Bbb C}{\Bbb P}^2$
to see the correspondence with its fuzzy counterpart.
The homogeneous coordinates $(z_1,z_2,z_3)$ on ${\Bbb C}{\Bbb P}^2$
are subject to the equivalence relations
\eqn\equivCPtwo{
(z_1,z_2,z_3) \sim (\lambda z_1, \lambda z_2, \lambda z_3)
}
with $\lambda \in {\Bbb C}^*$.
We can take the representative coordinates
$(\Phi_1,\Phi_2,\Phi_3)$
from the equivalent class which satisfy 
\eqn\commr{
|\Phi_1|^2 + |\Phi_2|^2 + |\Phi_3|^2 = r^2
}
for some fixed real number $r$.
The generalization of \commr\ 
to fuzzy ${\Bbb C}{\Bbb P}^2$
with the non-commutative coordinates $a_i$
corresponding to $\Phi_i$ $(i=1,2,3)$
would be
\eqn\fuzzyN{
a_1^\dagger a_1 + a_2^\dagger a_2 + a_3^\dagger a_3 \Bigr|_{{\cal F}^{(3)}_N}
= m_1+m_2+m_3 = N
}
with fixed $N$,
which appears in the definition 
\VNthree.
In the commutative case, we still have remaining equivalence relation
through the phase:
\eqn\equivCPtwoPhase{
(\Phi_1,\Phi_2,\Phi_3) \sim (e^{i \theta} \Phi_1, e^{i \theta} \Phi_2, e^{i \theta} \Phi_3)
}
with real number $\theta$.
The non-commutative coordinates
on fuzzy ${\Bbb C}{\Bbb P}^2$
should be also identified through such equivalence relation.
In our setting, 
this phase identification is provided by
the $U(1)$ subgroups of the gauge groups at the nodes,
as we will explain shortly.

From \dinFNthree,
$n_{1,2,3}$ in \ninN\ are given by
\eqn\nithree{
n_1=\hf(N+1)(N+2),\quad
n_2=\hf(N+2)(N+3),\quad
n_3=\hf(N+3)(N+4).
}
With this choice, $A_i$'s and $B_i$'s are interpreted as
the linear maps between the Fock spaces:
\eqn\ABasVNmap{
{\cal F}^{(3)}_{N+2}~\uerel{B_i}{\longrightarrow}~{\cal F}^{(3)}_{N+1}~
\uerel{A_i}{\longrightarrow}~{\cal F}^{(3)}_{N}~.
}
One obvious solution of \reducedABeq\ is obtained by setting
$A_i$ and $B_i$ to be equal to
the annihilation operator $a_i$ up to
complex number coefficients $c,\til{c}$:
\eqn\ABsolaa{
A_i=ca_i\Big|_{{\cal F}^{(3)}_{N+1}},\quad B_i=\til{c}a_i\Big|_{{\cal F}^{(3)}_{N+2}}~.
}
Then the F-term condition in \reducedABeq\ is automatically satisfied.
The remaining D-term condition reads
\eqn\Dcondinca{\eqalign{
|c|^2\sum_{i=1}^3a_i a_i^\dag&=\zeta_1{\bf 1}_{{\cal F}^{(3)}_{N}}\cr
\sum_{i=1}^3(|\til{c}|^2a_ia_i^\dag-|c|^2a_i^\dag a_i)&=\zeta_2{\bf 1}_{{\cal F}^{(3)}_{N+1}}\cr
-|\til{c}|^2\sum_{i=1}^3a_i^\dag a_i&=\zeta_3{\bf 1}_{{\cal F}^{(3)}_{N+2}}~.
}}
Using the relation
\eqn\aidagthree{
\sum_{i=1}^3a_ia_i^\dag=3+\sum_{i=1}^3a_i^\dag a_i~,
}
we find that
$c,\til{c}$ and the FI-parameters are related as
\eqn\zetainc{
\zeta_1=|c|^2(N+3),\quad
\zeta_2=|\til{c}|^2(N+4)-|c|^2(N+1),\quad
\zeta_3=-|\til{c}|^2(N+2).
}
One can check that the above $\zeta_r$ satisfy
\eqn\sumFIthree{
\sum_{r=1}^3n_r\zeta_r=0~,
}
which means that the overall $U(1)$ of the gauge group $\prod_r U(n_r)$ 
is decoupled.
On the other hand,
if we parametrize the $U(1)$ subgroup of $U(n_r)$
gauge group at the node $r$ by $e^{i \theta_r}$ $(r=1,2,3)$,
we obtain following gauge equivalence relations:
\eqn\phase{
A_i \sim e^{i (\theta_3-\theta_2)} A_i ,
\quad
B_i \sim e^{ (\theta_2-\theta_1)} B_i 
\quad (i=1,2,3).
}
Since the overall $U(1)$ phase decouples,
we have two independent $U(1)$ phases left,
which we can choose as $\theta_3-\theta_2$ 
and $\theta_2 - \theta_1$.
\phase\ correspond to \equivCPtwoPhase.
More precisely, there are two coordinate system
$A_i$ and $B_i$, and each of them has
one equivalence relation by the $U(1)$ phase.
Together 
with the restriction $m_1+m_2+m_3 = N$ in \VNthree,
our construction of fuzzy ${\Bbb C}{\Bbb P}^2$
can be regarded as a non-commutative version
of the geometric visualization of ${\Bbb C}{\Bbb P}^2$
from the toric diagram explained in \IqbalYE.
To add some words for explanation,
commutative ${\Bbb C}{\Bbb P}^2$ is described
as $S^1 \times S^1$ fibration over the base
represented by the edges and the inside of the triangle
of the toric diagram,
where $S^1 \times S^1$ fiber 
is given by the relative phases of $\Phi_i$.
The fiber degenerates at the edges of the triangle.
In our fuzzy ${\Bbb C}{\Bbb P}^2$, 
what is ``quantized'' is this base space,
which is given by a finite set of points as above.\foot{%
A construction of fuzzy toric geometries 
with a different procedure for quantizing
the toric base 
has been investigated in \SaemannGF.}

The expected dimension of the solution space of \reducedABeq\ is given by 
\eqn\dmoduli{
d=3n_1n_2+3n_2n_3-(n_1^2+n_2^2+n_3^2-1)-3n_1n_3~.
}
This can be understood as follows \DouglasQW.
The first and the second terms represent
the number of chiral multiplets $A_i,B_i$, and the 
middle term in the parenthesis is the dimension of the gauge
group $[\prod_r U(n_r) ]/U(1)$, and finally the last term in
\dmoduli\ is the number of F-term relations $A_iB_j=A_jB_i$.
For our choice of $n_{1,2,3}$ in \ninN, we find
$d=0$. Therefore, our solution corresponds to the ``exceptional
bundle'' on ${\Bbb C}{\Bbb P}^2$ \DouglasQW. 

Let us consider the D-brane charges carried by our solution.
As reviewed in appendix A,
the D-brane charge is related to the rank of gauge
group $n_{1,2,3}$ by\foot{In Fig. 1, 
we have reversed the direction of arrows compared to the
diagram in \DouglasQW, in order that $A_i,B_i$ 
correspond to the annihilation operator $a_i$,
not the creation operator $a^\dag_i$. Therefore,
$X_i,Y_i$ in 
\DouglasQW\ correspond to our $A_i^\dag,B_i^\dag$,
respectively. We have also changed the
overall sign of the central charge from 
\DouglasQW\ so that our solution represents
a D4-brane instead of a $\b{\rm D4}$-brane.
}
\eqn\QDcharge{
Q_{{\rm D}4}=n_1-2n_2+n_3,\quad
Q_{{\rm D}2}=n_2-n_1,\quad
Q_{{\rm D}0}={n_1+n_2\o2}~.
} 
From \ninN, we find that the D-brane charges
carried by our solution are given by
\eqn\QDsolone{
Q_{{\rm D}4}=1,\quad
Q_{{\rm D}2}=N+2,\quad
Q_{{\rm D}0}=\hf(N+2)^2~,
}
and the BPS central charge is
\eqn\chEinom{
Z=Q_{{\rm D}4}+Q_{{\rm D}2}\om+Q_{{\rm D}0}\om^2=e^{(N+2)\om}~,
}
where $\om$ denotes the generator of $H^2({\Bbb C}{\Bbb P}^2)$.
From \QDcharge\ we see that
our solution represents a single D4-brane wrapped on ${\Bbb C}{\Bbb P}^2$
with $N+2$ unit of magnetic flux threading its worldvolume
\eqn\FonDseven{
{F\o 2\pi}=(N+2)\om~.
}
Due to this magnetic flux, the worldvolume theory on the D4-brane
becomes non-commutative \SeibergVS\ with the non-commutativity 
parameter
\eqn\noncomth{
\th=F^{-1}\sim{1\o N+2}~.
}

We can also construct a solution describing $k$ D4-branes wrapping
${\Bbb C}{\Bbb P}^2$
by tensoring the Chan-Paton spaces in 
\ninN\ with ${\Bbb C}^k$:
\eqn\kDseven{
{\Bbb C}^{n_1}={\cal F}^{(3)}_{N}\tens{\Bbb C}^k,\quad
{\Bbb C}^{n_2}={\cal F}^{(3)}_{N+1}\tens{\Bbb C}^k,\quad
{\Bbb C}^{n_3}={\cal F}^{(3)}_{N+2}\tens{\Bbb C}^k~,
}
and generalizing
the single D4-brane solution \ABsolaa\ as
\eqn\ABksol{
A_i=ca_i\tens {\bf 1}_k,\quad
B_i=\til{c}a_i\tens {\bf 1}_k\quad(i=1,2,3)~.
}
In particular, by taking
$k=5$ we can construct 
a $U(5)$ gauge theory living on
D7-branes, which may have useful applications 
in the model building in F(uzz) theory.

\subsec{Fuzzy ${\Bbb C}{\Bbb P}^2$}
In this subsection, we will
show that our solution \ABsolaa\ represents a BPS D4-brane wrapped on
fuzzy ${\Bbb C}{\Bbb P}^2$.
One can repeat the analysis of fuzzy ${\Bbb C}{\Bbb P}^1$
in the previous section
by introducing the $su(3)$ generators
\eqn\Jathree{
J^a=a^\dag_iT^a_{ij}a_j~\quad(a=1,...,8)~,
}
where $T^a$'s are the generators of $su(3)$ in the fundamental representation.
The commutation relations $[T^a,T^b]=if^{abc}T^c$ and 
$[a_i,a_j^\dag]=\cob_{ij}$
imply
$[J^a,J^b]=if^{abc}J^c$. 
By restricting the action of $J^a$
to the finite dimensional space ${\cal F}^{(3)}_N$,
this construction leads to the fuzzy ${\Bbb C}{\Bbb P}^2$ 
\refs{\AlexanianQJ,\CarowWatamuraCT}.

More generally, we can consider $n$ independent oscillators
$[a_i,a_j^\dag]=\cob_{ij}~(i,j=1,\cdots,n)$,
and construct the $su(n)$ generators as $J^a=a^\dag_iT^a_{ij}a_j$.
Using the normalization $\Tr(T^aT^b)=\hf\cob^{ab}$
and the identity
\eqn\TaFierz{
T^a_{ij}T^a_{kl}=
\hf\cob_{il}\cob_{jk}-{1\o2n}\cob_{ij}\cob_{kl}~,
}
one can show that $J^a=a^\dag_iT^a_{ij}a_j$ satisfy
\eqn\Jarel{\eqalign{
J^aJ^a&={n-1\o2n}\h{N}(\h{N}+n)~,\cr
d_{abc}J^bJ^c&={n-2\o2n}(2\h{N}+n)J^a~.
}}
Here $d_{abc}=2\Tr (T^a\{T^b,T^c\})$ denotes the totally symmetric
tensor and $\h{N}=\sum_ia^\dag_ia_i$ is the total number
of oscillators.

We can define
the fuzzy ${\Bbb C}{\Bbb P}^2$ by
\eqn\deffuzzCPtwo{
{\Bbb C}{\Bbb P}^2_N={\rm End}({\cal F}^{(3)}_N)={\rm Mat}
\lf(\hf(N+1)(N+2),{\Bbb C}\ri)
}
and the generators of fuzzy ${\Bbb C}{\Bbb P}^2_N$
algebra is $J^a|_{{\cal F}^{(3)}_N}$.
The relations satisfied by these generators are
found by setting $n=3$ in \Jarel:
\eqn\relJaCPtwo{
J^aJ^a={1\o3}N(N+3),\quad
d_{abc}J^bJ^c={1\o6}(2N+3)J^a~.
} 
Note that the value of $J^aJ^a$ in \relJaCPtwo\
is the quadratic Casimir of the representation
${\rm Sym}^N{\bf 3}=[N,0]$ of $SU(3)$.

In 
the commutative language,
our solution \ABsolaa\ corresponds  to the line bundle
${\cal O}_{{\Bbb P}^2}(N+2)$,
whose global sections are the degree $N+2$ polynomials.
In the non-commutative setting,
these sections correspond to the states in ${\cal F}_{N+2}^{(3)}$,
which in turn is the largest space ${\Bbb C}^{n_3}$ in our
quiver \ninN. 

\newsec{KK Spectrum on (Fuzzy) ${\Bbb C}{\Bbb P}^2$}
In this section we study the KK spectrum on fuzzy ${\Bbb C}{\Bbb P}^2$.
We first review the spectrum of the scalar Laplacian on commutative 
${\Bbb C}{\Bbb P}^2$ and  non-commutative 
${\Bbb C}{\Bbb P}^2$. Next we study the
fluctuation spectrum of the gauge fields around our classical solution, and
we will find that this spectrum agrees with the
known eigenvalues of the scalar Laplacian on ${\Bbb C}{\Bbb P}^2$
for the low-lying modes.

\subsec{Spectrum of the Scalar Laplacian on Commutative ${\Bbb C}{\Bbb P}^2$}
Let us first recall the spectrum of the scalar Laplacian
on commutative ${\Bbb C}{\Bbb P}^2$ \Ikeda.
Since ${\Bbb C}{\Bbb P}^2$ is 
a symmetric space of the form
\eqn\PtwoasGH{
{\Bbb C}{\Bbb P}^2=SU(3)/S(U(2)\times U(1))
}
the Laplace equation $\lap\phi=\la\phi$ for the scalar Laplacian
\eqn\scalLap{
\lap=-{1\o\rt{g}}\del_\mu(g^{\mu\nu}\rt{g}\del_\nu)
}
can be solved by a group theoretical method.
The Fubini-Study metric of ${\Bbb C}{\Bbb P}^2$ is
\eqn\KFS{
g_{i\bar{j}}=\del_i\del_{\bar{j}}K,\quad K={1\o M_{\rm KK}^2}\log(|z_0|^2+|z_1|^2+|z_2|^2)~,
}
where $[z_0:z_1:z_2]$ are the homogeneous coordinates of
${\Bbb C}{\Bbb P}^2$ and $1/M_{\rm KK}$ 
sets the length scale of ${\Bbb C}{\Bbb P}^2$.
In terms of the inhomogeneous coordinates $w_i={z_i\o z_0}~(i=1,2)$ on the patch
$z_0\not=0$, the Laplacian is written as
\eqn\lapinwi{
\lap=-4M_{\rm KK}^2(1+|w_1|^2+|w_2|^2)\left[\sum_{i=1}^2{\del^2\o \del w_i\del\b{w_i}}
+\sum_{i,j=1}^2w_i{\del\o\del w_i}\b{w_j}{\del\o\del\b{w_j}}\right]~.
} 
The eigenvalues of the scalar Laplacian are given by
\eqn\lapeigenv{
\la_m=4M_{\rm KK}^2m(m+2),\quad(m=0,1,2,\cdots)~,
}
and the multiplicity of the $m$-th eigenvalue is
\eqn\multmum{
\mu_m=(m+1)^3~.
}
This can be understood from the fact that the eigenspace associated with
the $m$-th eigenvalue $\la_m$ 
transforms as the irreducible representation of $SU(3)$
with Dynkin label $[m,m]$. The dimension of the
$[m,m]$ representation is $\mu_m$, and the eigenvalue
$\la_m$ is proportional to the quadratic Casimir of
the $[m,m]$ representation. 

This eigenspace of the Laplacian is related to
the space of homogeneous polynomials in the following way \BraunJP. 
The space ${\Bbb C}[z_i]_k$ of degree $k$ polynomials in $z_0,z_1,z_2$ 
and its complex conjugate ${\Bbb C}[\b{z_i}]_k$ transform
under $SU(3)$
as ${\rm Sym}^k{\bf 3}$ and ${\rm Sym}^k\b{{\bf 3}}$, respectively.
Next we introduce
the space of functions on ${\Bbb C}{\Bbb P}^2$
\eqn\spaceFk{
{\cal P}_k={{\Bbb C}[z_0,z_1,z_2]_k\otimes {\Bbb C}[\b{z_0},\b{z_1},\b{z_2}]_k\o (|z_0|^2+|z_1|^2+|z_2|^2)^k}
}
whose dimension is
\eqn\dimFk{
{\rm dim}{\cal P}_k=\pmatrix{k+2\cr k}^2~.
}
From the decomposition of $SU(3)$ representation
\eqn\symtens{
{\rm Sym}^k{\bf 3}\otimes {\rm Sym}^k\b{{\bf 3}}
=\bigoplus_{m=0}^k[m,m]~,
}
we find that the multiplicity of the $m$-th eigenvalue $\la_m$ is
given by
\eqn\muminFm{
\mu_m={\rm dim}{\cal P}_m-{\rm dim}{\cal P}_{m-1}
=\pmatrix{m+2\cr m}^2-\pmatrix{m+1\cr m-1}^2=(m+1)^3~.
}
Clearly, this agrees with \multmum.

\subsec{Spectrum of the Scalar Laplacian on Fuzzy ${\Bbb C}{\Bbb P}^2$}
As discussed in \GrosseCI, the spectrum of the Laplacian on fuzzy
${\Bbb C}{\Bbb P}^2_N$ is truncated by the total occupation number $N$:
\eqn\spectrum{
\la_m=4M_{\rm KK}^2m(m+2),\quad(m=0,1,\cdots,N)~.
}
This is because the space of homogeneous functions
${\cal P}_m$ in the commutative case
is replaced by the space of operators acting on the Fock space
of three independent oscillators ${\cal F}_N^{(3)}$ \VNthree.
For instance, the monomial
\eqn\monoz{
{z_0^{m_1}z_1^{m_2}z_2^{m_3}\b{z_0}^{m'_1}\b{z_1}^{m'_2}\b{z_2}^{m'_3} 
\o (|z_0|^2+|z_1|^2+|z_2|^2)^m}~,
\qquad\Big(\sum_{i=1}^3m_i=\sum_{i=1}^3m'_i=m\Big)
}
corresponds to the operator acting on ${\cal F}^{(3)}_N$
\eqn\monoonFN{
{1\o N^m}(a^\dag_1)^{m'_1}(a^\dag_2)^{m'_2}(a^\dag_3)^{m'_3}
(a_1)^{m_1}(a_2)^{m_2}(a_3)^{m_3}~.
}
Here we have used the usual normal ordering prescription
to go from the commutative expression \monoz\ to the operator
\monoonFN.
The upper bound for the harmonics $m\leq N$ comes from
the fact that the operator in \monoonFN\ annihilates all states
in ${\cal F}_N^{(3)}$
if $m$ exceeds $N$.  We can also check that
the total number of eigen-modes
agrees with the dimension of
the algebra ${\Bbb C}{\Bbb P}^2_N={\rm End}({\cal F}_N^{(3)})$:
\eqn\summum{
\sum_{m=0}^N\mu_m=\sum_{m=0}^N(m+1)^3
=\left\{\hf (N+1)(N+2)\right\}^2=\big({\rm dim}{\cal F}_N^{(3)}\big)^2
}
As discussed in \HeckmanPV,
the truncation of the KK spectrum in the F(uzz) theory may have
interesting consequences 
in the low energy physics.

\subsec{KK Reduction of the Gauge Fields}
The fluctuation of gauge fields around the fuzzy ${\Bbb C}{\Bbb P}^1$
solution in the $A_1$ quiver theory 
was studied in \refs{\BerensteinRI,\MaldacenaMW}.
Here we repeat similar analysis for fuzzy ${\Bbb C}{\Bbb P}^2$.
The KK mass spectrum of the gauge fields
is obtained from the kinetic terms of
$A_i$ and $B_i$:
\eqn\kinAi{\eqalign{
&{\cal L}=\sum_{i=1}^3\Tr \Big(|D_t A_i|^2+|D_t B_i|^2\Big),\cr
&D_t A_i=\del_t A_i+i(W_t^1 A_i-A_iW_t^2),\cr
&D_t B_i=\del_t B_i+i(W_t^2 B_i-B_iW_t^3),
}}
where $W_t^{1,2,3}$ denotes the 1-dimensional
gauge field 
at the nodes 1, 2, and 3. 
From our interpretation  
of the solution as a D4-brane,
we expect to obtain
a single KK tower for a single gauge field
living on the worldvolume of the single D4-brane,
given by the eigenvalues of the 
scalar Laplacian on fuzzy ${\Bbb C}{\Bbb P}^2$ \lapeigenv. 
We will see below that this is indeed true
for the low-lying excitations in the large $N$ limit.

Expanding \kinAi\ around the classical solution
\ABsolaa,
the mass term for the gauge field becomes
\eqn\Wmumassterm{\eqalign{
{\cal L}_{\rm mass}&=|c|^2\sum_{i=1}^3\Tr\Big[a_ia_i^\dag(W_t^1)^2
+a_i^\dag a_i(W_t^2)^2-a_i^\dag W_t^1a_i W_t^2
-W_t^1a_iW_t^2a_i^\dag\Big]\cr
&+|\til{c}|^2\sum_{i=1}^3\Tr\Big[a_ia_i^\dag(W_t^2)^2
+a_i^\dag a_i(W_t^3)^2-a_i^\dag W_t^2a_i W_t^3
-W_t^2a_iW_t^3a_i^\dag\Big]~.
}}
As discussed in \refs{\BerensteinRI,\MaldacenaMW},
since the eigenvalue is common for all scalar harmonics
belonging to the same representation $[m,m]$ of $SU(3)$, 
it is sufficient to consider one particular mode
\eqn\phim{
\phi_m=(a_1^\dag a_2)^m.
}
We expand the gauge fields as
\eqn\Wmuharm{
W_t^r=g{W_t^{r,m}\cdot(\phi_{r,m}+\phi_{r,m}^\dag)\o \rt{\Tr(\phi_{r,m}+\phi_{r,m}^\dag)^2}},\qquad
(r=1,2,3)
}
where $\phi_{r,m}$ denotes the restriction of 
$\phi_m$ to the $r$-th Chan-Paton space ${\Bbb C}^{n_r}$ in \ninN,
and $g$ denotes the gauge coupling. Here we have 
assumed for simplicity that
the gauge couplings of all three gauge groups $U(n_r)$ have
a common value $g$.
The normalization factor
$g/\rt{\Tr(\phi_{r,m}+\phi_{r,m}^\dag)^2}$
in \Wmuharm\ is included 
so that the kinetic term of
$W_t^{r,m}$ becomes canonical \BerensteinRI.

The diagonal terms of the mass matrix in \Wmumassterm\
are found by using \aidagthree.
In order to evaluate the mixing terms in \Wmumassterm,
we first notice the relation
\eqn\aisandW{\eqalign{
\sum_{i=1}^3 a_i^\dag\phi_m a_i&=(\h{N}-m)\phi_m\cr
\sum_{i=1}^3 a_i \phi_m a_i^\dag&=(\h{N}+3+m)\phi_m~,
}}
where $\h{N}=\sum_{i=1}^3 a_i^\dag a_i$ is the total number operator.
From this relation, we obtain
\eqn\phiimasand{\eqalign{
\sum_{i=1}^3 a_i^\dag(\phi_{1,m}+\phi_{1,m}^\dag)a_i&=
(N+1-m)(\phi_{2,m}+\phi_{2,m}^\dag) \cr
\sum_{i=1}^3 a_i(\phi_{2,m}+\phi_{2,m}^\dag)a_i^\dag &=
(N+3+m)(\phi_{1,m}+\phi_{1,m}^\dag) \cr
\sum_{i=1}^3 a_i^\dag(\phi_{2,m}+\phi_{2,m}^\dag)a_i&=
(N+2-m)(\phi_{3,m}+\phi_{3,m}^\dag) \cr
\sum_{i=1}^3 a_i(\phi_{3,m}+\phi_{3,m}^\dag)a_i^\dag &=
(N+4+m)(\phi_{2,m}+\phi_{2,m}^\dag)~.
}}
Furthermore, by noticing that
\eqn\tracerel{\eqalign{
&\sum_{i=1}^3\Tr \,(\phi_{1,m}+\phi_{1,m}^\dag)a_i
(\phi_{2,m}+\phi_{2,m}^\dag)a_i^\dag
=(N+3+m)\Tr(\phi_{1,m}+\phi_{1,m}^\dag)^2\cr
=&\sum_{i=1}^3\Tr \,a_i^\dag(\phi_{1,m}+\phi_{1,m}^\dag)a_i
(\phi_{2,m}+\phi_{2,m}^\dag)=
(N+1-m)\Tr(\phi_{2,m}+\phi_{2,m}^\dag)^2~,
}}
we find the identity
\eqn\traceid{\eqalign{
&(N+3+m)\Tr(\phi_{1,m}+\phi_{1,m}^\dag)^2=
(N+1-m)\Tr(\phi_{2,m}+\phi_{2,m}^\dag)^2\cr
=&\rt{(N+3+m)(N+1-m)\Tr(\phi_{1,m}+\phi_{1,m}^\dag)^2
\Tr(\phi_{2,m}+\phi_{2,m}^\dag)^2}~.
}
}
Similarly, the mixing term of $W^2_t$ and $W^3_t$ becomes
\eqn\tracefortwothree{\eqalign{
&\sum_{i=1}^3\Tr \,a_i^\dag(\phi_{2,m}+\phi_{2,m}^\dag)a_i
(\phi_{3,m}+\phi_{3,m}^\dag)\cr
=&\sum_{i=1}^3\Tr \,(\phi_{2,m}+\phi_{2,m}^\dag)a_i
(\phi_{3,m}+\phi_{3,m}^\dag)a_i^\dag\cr
=&\rt{(N+4+m)(N+2-m)\Tr(\phi_{2,m}+\phi_{2,m}^\dag)^2
\Tr(\phi_{3,m}+\phi_{3,m}^\dag)^2}~.
}
}
Finally, we find that 
the mass matrix of $W_t^{1,2,3}$ in the representation
$[m,m]$ is given by
\eqn\massmatW{\eqalign{
{\cal M}=&|gc|^2\pmatrix{N+3&-\rt{(N+3)(N+1)-m(m+2)}&0\cr 
-\rt{(N+3)(N+1)-m(m+2)}& N+1&0\cr 
0&0& 0} \cr
+&
|g\til{c}|^2\pmatrix{0&0&0\cr 
0& N+4& -\rt{(N+4)(N+2)-m(m+2)}\cr 
0& -\rt{(N+4)(N+2)-m(m+2)}& N+2}.
}}

The KK spectrum is obtained by diagonalizing
this matrix. 
The characteristic equation $\det(\la-{\cal M})=0$
for the mass matrix reads
\eqn\Mcalpoly{\eqalign{
&\la^3-\Big\{|gc|^2(N+2)+|g\til{c}|^2(N+3)\Big\}
\Big\{2\la^2+|gc|^2|g\til{c}|^2m(m+2)\Big\}\cr
&\qquad\qquad+\la\Big\{m(m+2)(|gc|^4+|g\til{c}|^4)+2|gc|^2|g\til{c}|^2(n_1+n_2+n_3)\Big\}=0~,
}}
and the KK eigenvalues are given by the solution of this cubic
equation. Writing the exact solution of this characteristic equation is
not so illuminating, so we consider the
approximate behavior of the small $m$ eigenvalues 
in the large $N$ regime.
When $m\ll N$, the eigenvalues are found to be
\eqn\lamlargeN{\eqalign{
\la_{m,0}&\sim{|gc|^2(N+2)+|g\til{c}|^2(N+3)\o2(n_1+n_2+n_3)}m(m+2)~,\cr
\la_{m,\pm}&\sim
|gc|^2(N+2)+|g\til{c}|^2(N+3)\cr
&\quad~~~\pm\rt{\big\{|gc|^2(N+2)-|g\til{c}|^2(N+3)\big\}^2
+|gc|^2|g\til{c}|^2(N+1)(N+4)}~.
}}
We can see that $\la_{m,0}$ corresponds to the 
$m$-th eigenvalue of the scalar Laplacian on ${\Bbb C}{\Bbb P}^2$ \lapeigenv.
Therefore, the low-lying KK modes reproduce the known
spectrum of the scalar Laplacian on ${\Bbb C}{\Bbb P}^2$.
By comparing the commutative expression
$\la_m=4M_{\rm KK}^2m(m+2)$ in \lapeigenv\ and the eigenvalue
$\la_{m,0}$ in \lamlargeN,
we find that the KK mass scale for the compactification
on fuzzy ${\Bbb C}{\Bbb P}^2$
is given by
\eqn\cvsMKK{
M_{\rm KK}^2\sim ~{|gc|^2(N+2)+|g\til{c}|^2(N+3)\o n_1+n_2+n_3}~.
}
This KK mass scale behaves as $M_{\rm KK}^2\sim N^{-1}$ when $N\gg1$, 
which suggests that
our fuzzy ${\Bbb C}{\Bbb P}^2$
solution approaches a smooth large volume ${\Bbb C}{\Bbb P}^2$
in the large $N$ limit.
On the other hand, the interpretation of
other eigenvalues $\la_{m,\pm}$ in \lamlargeN\
is not so clear. As discussed in \MaldacenaMW,
$\la_{m,\pm}$ may be related to a UV effect
in the theory on fuzzy ${\Bbb C}{\Bbb P}^2$.
Note that $|gc|,|g\til{c}|$ in \cvsMKK\ have
the usual form of W-boson mass since the gauge symmetry
is completely Higgsed by the VEV of the scalar fields 
 $\bra A_i\ket= ca_i, \bra B_i\ket=\til{c}a_i$.

The expression of the mass matrix in \massmatW\ is valid for $m=0,\cdots,N$.
When $m=N+1$, we have $\phi_{1,m}=0$ and hence 
we have to diagonalize the mass matrix for
$W^{2,3}_t$.
Similarly, for $m=N+2$ we have $\phi_{1,m}=\phi_{2,m}=0$,
and the eigenvalue of $W_t^3$ is simply $|g\til{c}|^2(N+2)$.

\subsec{Mass Matrix for the Field $C_i$}
From the superpotential
$W=\vep^{ijk}\Tr(A_iB_jC_k)$,
we obtain the potential for the field $C_i$ as follows:
\eqn\pote{\eqalign{
V(C_i) &=
\left|
{\del W} \over {\del A_k}
\right|^2
+
\left|
{\del W} \over {\del B_k}
\right|^2 \cr
=& 
\Tr_{{\cal F}_{N+2}} 
(
C_i A_j A_j^{\dagger}  C_i^{\dagger} 
- C_i A_j A_i^{\dagger}  C_j^{\dagger} 
)
\cr
&+
 \Tr_{{\cal F}_{N}}  
(
C_i^{\dagger} B_j^{\dagger} B_j C_i 
- C_i^{\dagger} B_j^{\dagger} B_i C_j 
).
}
}
Expanding around the fuzzy ${\Bbb C}{\Bbb P}^2$ solution
\ABsolaa\
up to the quadratic order\foot{The D-term potential does not contain the term
quadratic in the fluctuation of the filed $C_i$ around the classical solution \ABsolaa.},
we obtain the mass term for the field $C_i$:
\eqn\quadratic{\eqalign{
&|c|^2 
\Tr_{{\cal F}_{N+2}} 
\lf\{
C_i   
(\delta^{ii'}\delta^{jj'} - \delta^{ij'}\delta^{ji'})
a_{j} a^{\dagger}_{j'} \Big|_{{\cal F}_{N}}
C_{i'}^{\dagger} 
\ri\}
\cr
+
&|\til{c}|^2
\Tr_{{\cal F}_{N}}  
\lf\{
C_{i'}^{\dagger} 
(\delta^{ii'}\delta^{jj'} - \delta^{i'j'} \delta^{ji} )
a^{\dagger}_{j} a_{j'} \Big|_{{\cal F}_{N+2}}
C_{i}
\ri\}~.
}
}
In the first line of \quadratic, 
we can rewrite as
\eqn\massprojA{\eqalign{
(\delta^{ii'}\delta^{jj'} - \delta^{ij'}\delta^{ji'})
a_{j} a^{\dagger}_{j'} \Big|_{{\cal F}_{N}}
&=
(N+3)
\delta^{ii'}
- {a_{i'} a^{\dagger}_i} \Big|_{{\cal F}_{N}}\cr
&= (N+2) 
\left( 
\delta^{ii'} - {{a^{\dagger}_i a_{i'}}\over {N+2}}
\right) \Bigl|_{{\cal F}_{N}}  .
}
}
Similarly, in the second line of \quadratic,
\eqn\massprojB{\eqalign{
(\delta^{ii'}\delta^{jj'} - \delta^{i'j'}\delta^{ji})
a^{\dagger}_{j} a_{j'} \bigl|_{{\cal F}_{N+2}}
&=
\delta^{ii'} (N+2) - a_{i}^{\dagger} a_{i'} \bigl|_{{\cal F}_{N+2}} 
\cr
&=
\delta^{ii'} (N+3)
\left(
\delta^{ii'} -
{ a_{i'} a_{i}^{\dagger} \over N+3}
\right)\Bigl|_{{\cal F}_{N+2}} .
}
}
We may write the mass term \quadratic\
in the following form: 
\eqn\DefMassMatrix{
C^i_{\vec{n}_1 \vec{n}_2}
 {\cal M}^{ij}_{(\vec{n}_1\vec{n}_2)(\vec{n}_3\vec{n}_4)}
C^{\dagger j}_{\vec{n}_3\vec{n}_4}  ,
}
where we have introduced a shorthand notation
\eqn\shorthand{
| \vn \rangle = | n_1, n_2, n_3 \rangle ,\quad
\langle \vn | = \langle n_1, n_2, n_3  | ,
}
and
\eqn\expandC{
C^i = C^i_{\vn_1\vn_2} | \vn_1 \rangle \langle \vn_2 | .
}
In \DefMassMatrix,
$|\vn_1 \rangle, |\vn_4 \rangle \in \F_{N+2}$
and 
$|\vn_2 \rangle,|\vn_3 \rangle \in \F_{N}$.
Using
\massprojA\ and \massprojB,
the mass matrix ${\cal M}$ is obtained as
\eqn\MassMatrix{\eqalign{
{\cal M}^{ij}_{(\vec{n}_1\vec{n}_2)(\vec{n}_3\vec{n}_4)}
=&
|c|^2 (N+2) (1-P_A)^{ij}_{(\vec{n}_1\vec{n}_2)(\vec{n}_3\vec{n}_4)} \cr 
&+
|\til{c}|^2 (N+3)(1-P_B)^{ij}_{(\vec{n}_1\vec{n}_2)(\vec{n}_3\vec{n}_4)},
}
}
where
\eqn\ProjA{
P^{ij}_{A(\vec{n}_1\vec{n}_2)(\vec{n}_3\vec{n}_4)}  
\equiv 
\left(
{a_i^{\dagger}a_j \over N+2} \Bigl|_{\F_N} 
\right)_{\vn_2 \vn_3} 
\left(
{\bf 1}_{\F_{N+2}} 
\right)_{\vn_1\vn_4} ,
}
\eqn\ProjB{
P^{ij}_{B(\vec{n}_1\vec{n}_2)(\vec{n}_3\vec{n}_4)}   
\equiv 
\left({\bf 1}_{\F_{N}}
\right)_{\vn_2\vn_3}
\left(  
{a_j a_i^{\dagger} \over N+3} \Bigl|_{\F_{N+2}}
\right)_{\vn_4\vn_1} ,
}
are projections.
Dimensions of these projections are given by
\eqn\dimProjA{\eqalign{
\dim P_A = \Tr P_A 
&= \Tr_{\F_N} {a_i^{\dagger}a_i \over N+2} \Tr_{\F_{N+2}} {\bf 1} 
\cr
&= 
{N \over N+2}
\dim \F_{N} 
\dim \F_{N+2} \cr
&= 
\dim \F_{N-1} \F_{N+2} ,
}
}
(here we assume $N \geq 2$), and
\eqn\dimProjB{\eqalign{
\dim P_B = \Tr P_B
&= 
\Tr_{\F_{N}} {\bf 1} \, \, 
\Tr_{\F_{N+2}} {a_i a_i^{\dagger} \over N+3} \cr
&= 
\dim \F_{N} \dim \F_{N+2} 
{N+5 \over N+3} \cr
&=
\dim \F_{N} \dim \F_{N+3}  . 
}
}
We can explicitly construct the basis 
of the projected space for each projection.
For the projection $P_A$, they are given by
\eqn\BasisProjA{
\left(a_j^{\dagger}\right)_{\vn_3 \vn} C^A_{\vn \vn_4} ~,
}
where
$|\vn_3 \rangle \in \F_N$, $|\vn \rangle \in \F_{N-1}$, $|\vn_4 \rangle \in \F_{N+2}$.
$C^A_{\vn \vn_2}$ is a 
$\dim \F_{N-1} \times \dim \F_{N+2}$ complex matrix,
so there are $\dim \F_{N-1} \times \dim \F_{N+2}$ independent such matrices,
in agreement with \dimProjA.\foot{The mass term 
\DefMassMatrix\
has a natural Hermitian structure.}
For the projection $P_B$, those are given by
\eqn\BasisProjB{
C^{B \dagger}_{\vn_3 \vn'} \left(a_j^{\dagger}\right)_{\vn' \vn_4}  .
}
where
$|\vn_3 \rangle \in \F_{N}$, 
$|\vn' \rangle \in \F_{N+3}$, $|\vn_4 \rangle \in \F_{N+2}$.
$C^{B \dagger}_{\vn_3 \vn'}$ is a 
$\dim \F_{N} \times \dim \F_{N+3}$ complex matrix,
so there are $\dim \F_{N} \times \dim \F_{N+3}$ independent such matrices,
in agreement with \dimProjB.

When $\til{c}=0$ (or ${c}=0$, respectively below), 
the mass is $|c|^2(N+2)$ ($|\til{c}|^2 (N+3)$)
for the components of $C_i$ projected by $1-P_A$ ($1-P_B$),
and zero for those projected by $P_A$ ($P_B$).
However,
since two projections $P_A$ and $P_B$ do not commute with each other 
i.e.
$[P_A,P_B] \ne 0$,
they cannot be simultaneously diagonalizable.
Therefore, diagonalization of the mass matrix
is a rather non-trivial problem 
when $c \til{c} \ne 0$.
We leave this problem
to the future.

\newsec{Fuzzy del Pezzo Surfaces from Quiver Gauge Theories}
In this section we will construct classical solutions
of the quiver gauge theories on the D0-branes probing 
del Pezzo singularities.
Our solutions describe 
D4-branes wrapped on fuzzy $dP_k$ for $k=1,2,3$.

\subsec{Fuzzy $dP_1$}
\fig{(a) The quiver diagram for the worldline theory
of D0-branes probing a $dP_1$ singularity.
(b) A classical solution describing
D4-branes wrapped on fuzzy $dP_1$ can be obtained
by setting $C_i=0$.
This corresponds to deleting the arrows between the node 0 and 
the node 3 
in the diagram (a). 
}{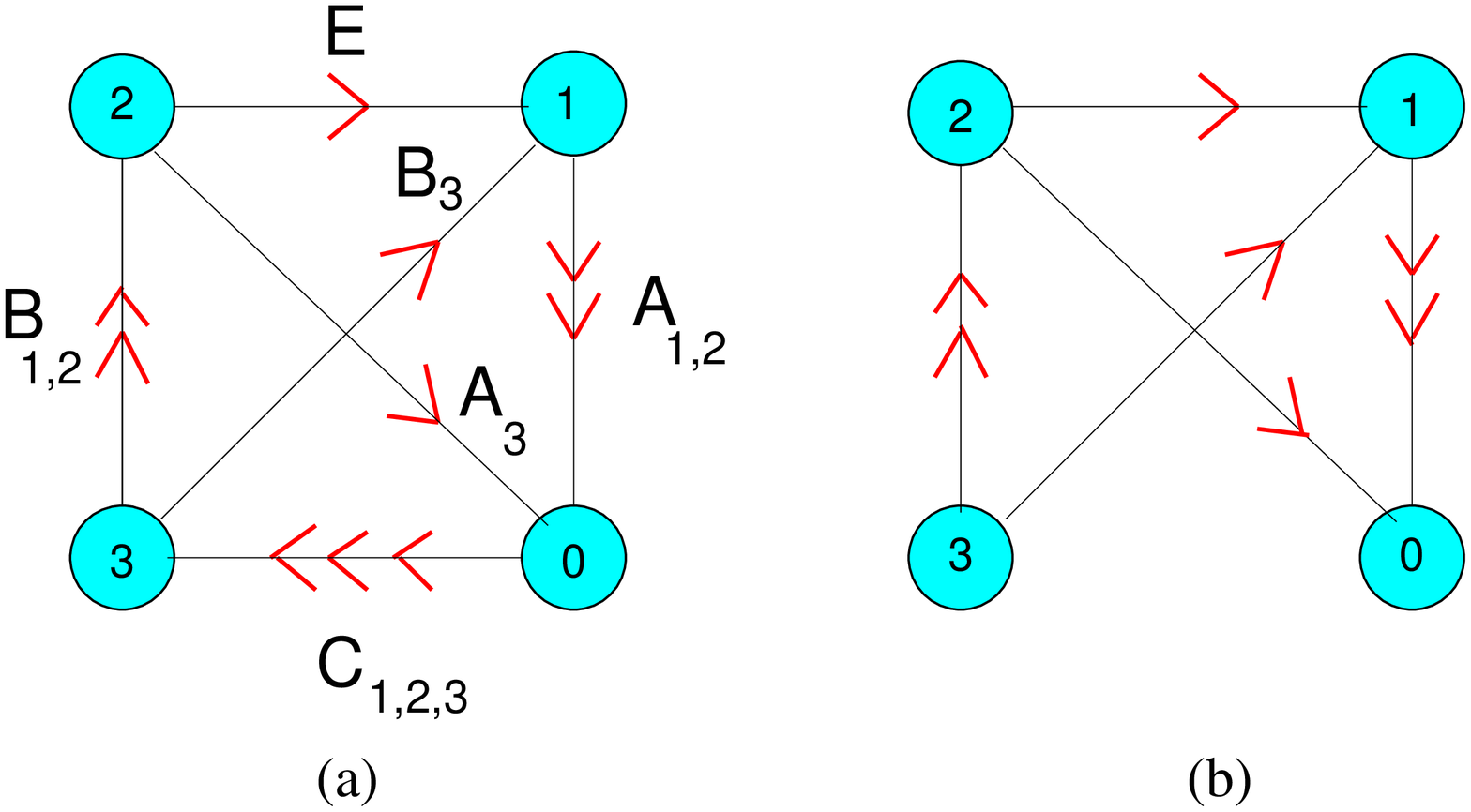}{7cm}
Let us first construct a classical solution
of quiver gauge theory for the D0-branes probing the $dP_1$ singularity.
The quiver diagram of this theory is depicted in Fig. 3a.
The superpotential of this quiver gauge theory is given by 
\refs{\FengMI,\BeasleyZP,\FrancoRJ,\KrippendorfHJ}
\eqn\supWdP{
W=\Tr\Big[A_1EB_2C_3-A_2EB_1C_3
+A_2B_3C_1-A_3B_2C_1
+A_3B_1C_2-A_1B_3C_2\Big]~.
}
We consider 
a classical solution with $C_i=0~(i=1,2,3)$.
This corresponds to deleting 
the arrows between the
node 0 and the node 3 (see Fig. 3b\foot{This is different
from the Beilinson quiver of $dP_1$ considered in \AspinwallVM, eq.(33).
However, as discussed around eq.(66) in \AspinwallVM,
we can transform the Beilinson quiver of $dP_1$  in \AspinwallVM\
into our Beilinson quiver in Fig. 3b by a series of mutations.

For further discussions on the Beilinson quiver in related contexts,
see \refs{\Bondal,\GovindarajanVI,\TomasielloYM,\HerzogQW,\HananyNM}.}).
After deleting those arrows, the remaining non-zero fields are
$A_i,B_i~(i=1,2,3,)$ and $E$. 
When $C_i=0$
the F-term condition for the other non-zero fields reads
\eqn\FforABE{
A_1EB_2=A_2EB_1,\quad
A_2B_3=A_3B_2,\quad
A_3B_1=A_1B_3~.
}
The D-term condition is 
\eqn\ABEDterm{\eqalign{
\zeta_0{\bf 1}_{V_0}&=A_1A_1^\dag+A_2A_2^\dag+A_3A_3^\dag\cr
\zeta_1{\bf 1}_{V_1}&=EE^\dag+B_3B_3^\dag-
A_1^\dag A_1-A_2^\dag A_2\cr
\zeta_2{\bf 1}_{V_2}&=B_1B_1^\dag+B_2B_2^\dag-E^\dag E-A_3^\dag A_3\cr
\zeta_3{\bf 1}_{V_3}&=-B_1^\dag B_1-B_2^\dag B_2-B_3^\dag B_3~.
}}
One obvious solution to the F-term \FforABE\ and the D-term
conditions \ABEDterm\ is
obtained by setting 
\eqn\BisoldP{
A_i=ca_i,\quad B_i=\til{c}a_i\quad(i=1,2,3),
\quad E=\rt{|c|^2+|\til{c}|^2}a_4,
} 
where
$a_i$'s are four independent
oscillators and 
$c,\til{c}$ are complex numbers.
In order to satisfy the D-term condition
\ABEDterm, we take the Chan-Paton spaces as
\eqn\ViFNM{
V_3={\cal F}_{N+2,M+1},\quad
V_2={\cal F}_{N+1,M+1},\quad
V_1={\cal F}_{N+1,M},\quad
V_0={\cal F}_{N,M},
}
where 
$V_r$ $(r=0,1,2,3)$ represents 
the Chan-Paton space at the node $r$ and
the space ${\cal F}_{N,M}$ is defined by
\eqn\defFMN{
{\cal F}_{N,M}=\Big\{|m_1,m_2,m_3,m_4\ket,\quad
\sum_{i=1}^3m_i=N,\quad m_3+m_4=M\Big\}~.
}
Then the FI parameter is quantized as
\eqn\FINM{\eqalign{
\zeta_0&=|c|^2(N+3)\cr
\zeta_1&=-|c|^2(N-M)+|\til{c}|^2(M+2)\cr
\zeta_2&=-|c|^2(M+1)+|\til{c}|^2(N-M+2)\cr
\zeta_3&=-|\til{c}|^2(N+2)~.
}}
The dimension of the space ${\cal F}_{N,M}$
can be counted from the quantized toric diagram
of the fuzzy $dP_1$ (Fig. 4).
It is given by the total number of the black points:
\eqn\dinFNM{\eqalign{
{\rm dim}\,{\cal F}_{N,M}
&=\hf(N+1)(N+2) - \hf(N-M)(N-M+1) \cr
&=\hf(M+1)(2N+2-M)~.
}
}
%
\fig{``Quantized'' toric diagram of fuzzy $dP_1$.
The black points are the points in the 
base space of fuzzy $dP_1$ when it is looked 
as a torus fibered over the base space.
Compared with the toric diagram of fuzzy ${\Bbb C}{\Bbb P}^2$ 
Fig. 2, white points in this figure are removed,
which is regarded as a fuzzy version of the blow-up at a point
$z_1 = z_2 =0$.
}{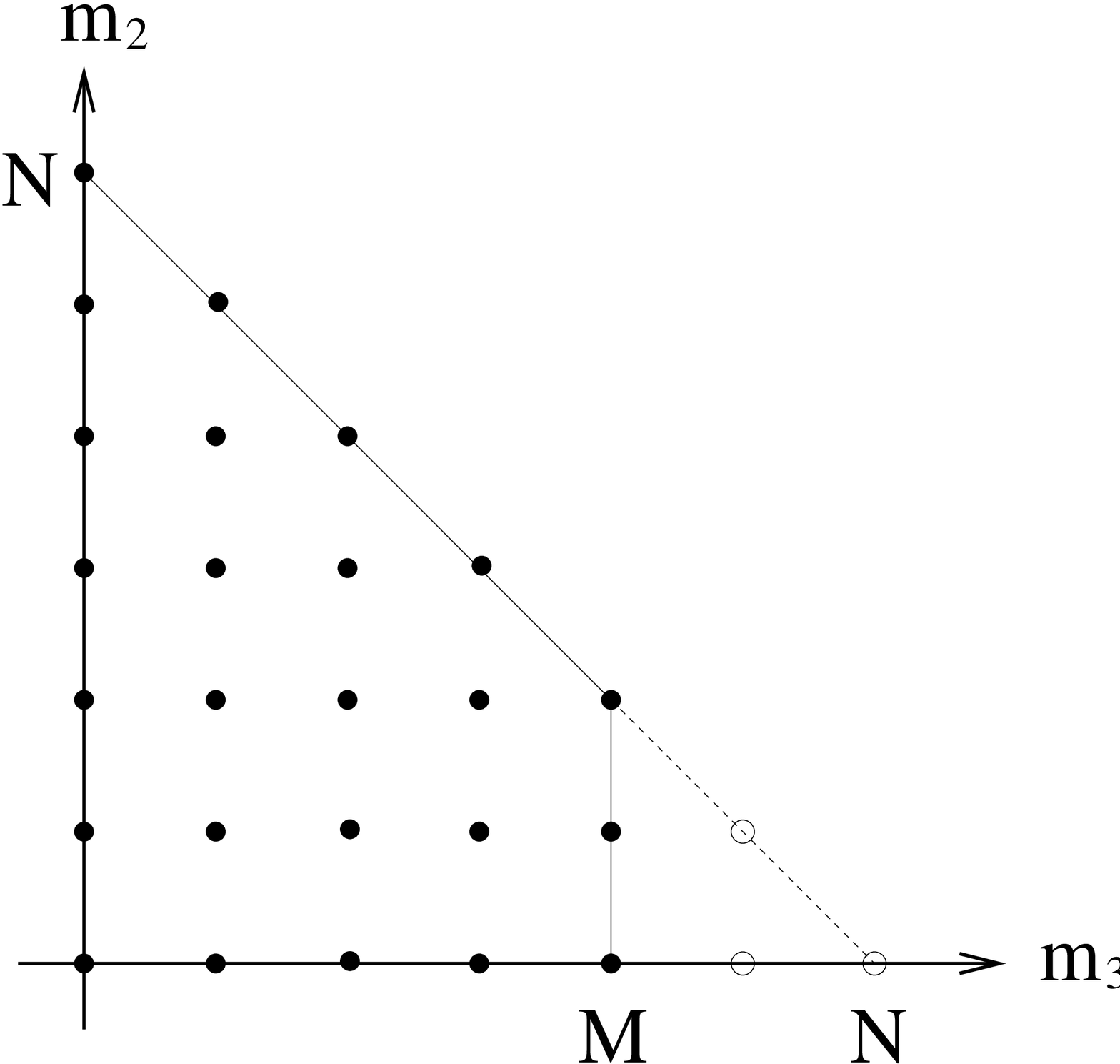}{5cm}
\noindent Using \dinFNM,
one can show that
\eqn\Finisum{
\sum_{r=0}^3\zeta_r{\rm dim}\,V_r=0~,
}
which means that the overall $U(1)$ subgroup of the gauge group is decoupled.
Let us represent the $U(1)$ subgroup of the $U(n_r)$ gauge group 
at the node $r$ by $e^{i \theta_r}$ $(r=0,1,2,3)$.
Since the overall $U(1)$ phase decouples,
we have three $U(1)$ phases left.
We have gauge equivalence relations
\eqn\phaseDPone{\eqalign{
A_{1,2} \sim e^{i (\theta_1-\theta_0)} A_{1,2}, &
\quad A_3 \sim e^{i(\theta_2-\theta_0)} A_3
= e^{i((\theta_2-\theta_1)+(\theta_1-\theta_0))} A_3, \cr
B_{1,2} \sim e^{i (\theta_3-\theta_2)} B_{1,2}, &
\quad B_3 \sim e^{i(\theta_3-\theta_1)} B_3
= e^{i((\theta_3-\theta_2) + (\theta_2-\theta_1))} B_3, \cr
E \sim  e^{i(\theta_2-\theta_1)} E~, &
}
}
thus we can choose $\theta_1-\theta_0$,
$\theta_3 - \theta_2$, $\theta_2-\theta_1$
as three independent $U(1)$ phases.
We observe that \phaseDPone\ corresponds
to eq.(33) of \IqbalYE,
while \defFMN\ corresponds
to eq.(32) of \IqbalYE.
On the other hand,
$m_1+m_2+m_3=N$ and $m_3+m_4=M$ in \defFMN\
corresponds to eq.(32) of \IqbalYE.
Thus, our classical solution \BisoldP\
can be regarded as a fuzzy version 
of the $dP_1$ surface.
Let us have a closer look on our fuzzy $dP_1$.
Recall that the toric diagram
of ${\Bbb C}{\Bbb P}^2$ (Fig. 2).
Since blowing up at a point on ${\Bbb C}{\Bbb P}^2$
corresponds 
to replacing a vertex of the triangle by a line segment,
the blowing-up procedure in the fuzzy setup can be realized 
by removing some of the states in
the Fock space ${\cal F}_N^{(3)}$. 
This is indeed the case for our
space ${\cal F}_{N,M}={\cal F}_{N,M}(dP_1)$.
If we suppress the dependence on the fourth 
oscillator, the condition $m_3+m_4=M$
implies the upper bound on $m_3$:
\eqn\FNdP{
{\cal F}_{N,M}(dP_1)=\left\{|m_1,m_2,m_3\ket,\quad
\sum_{i=1}^3m_i=N,\quad m_3\leq M\right\}.
}
Hence the states with $m_3>M$ are removed from
${\cal F}_N^{(3)}$ as indicated by 
by the white points in Fig. 4.
The vertex at $|0,0,N\ket$ is 
replaced by fuzzy 
${\Bbb C}{\Bbb P}^1$
\eqn\excPone{
{\cal F}_{N-M}({\Bbb C}{\Bbb P}^1)=\Big\{|m_1,m_2,M\ket,~m_1+m_2=N-M\Big\}~,
}
which corresponds to the 
exceptional curve appeared in the
blow-up of ${\Bbb C}{\Bbb P}^2$ at a point.

As discussed in \HeckmanPV, by sandwiching between the projection operator
\eqn\projectpi{
{\cal P}_{dP_1}:~{\cal F}_N^{(3)}={\cal F}_N({\Bbb C}{\Bbb P}^2)~\riya
~{\cal F}_{N,M}(dP_1)
}
we can compute various quantities such as Yukawa couplings on
fuzzy $dP_1$ from the knowledge of the parent fuzzy ${\Bbb C}{\Bbb P}^2$.

\subsec{Fuzzy $dP_2$}
\fig{(a) The quiver diagram for the worldline theory
of D0-branes probing the $dP_2$ singularity.
(b) A classical solution describing
D4-branes wrapped on fuzzy $dP_2$ can be obtained
by setting $B_i=C_i=0$.
}{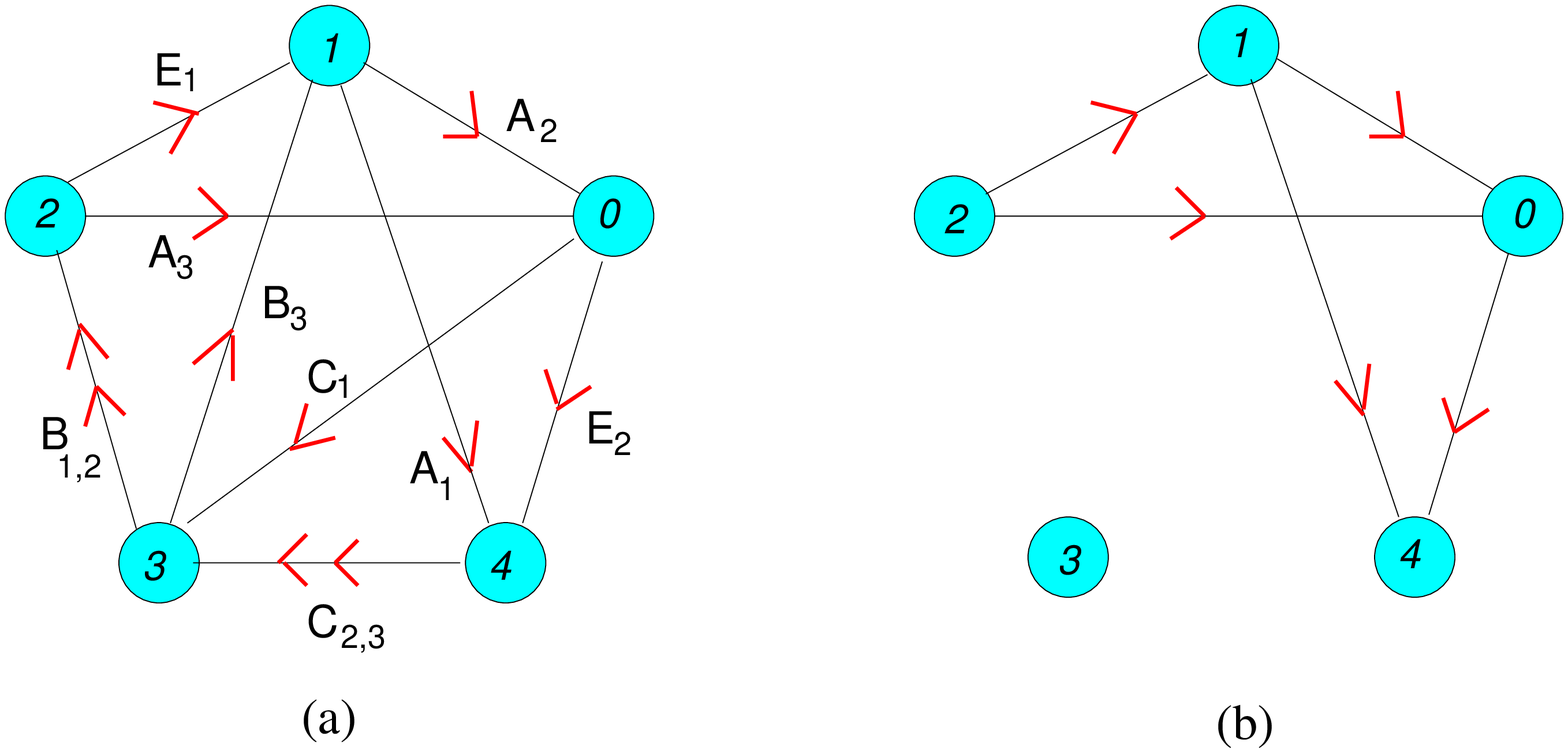}{7cm}
We can construct a classical solution of the
$dP_2$ quiver gauge theory
in a similar manner as in the $dP_1$ case in the previous subsection.
The quiver diagram for the
$dP_2$ singularity is depicted in
Fig. 5a. The superpotential is
given by \refs{\FengMI,\BeasleyZP,\FrancoRJ,\KrippendorfHJ}
\eqn\WdPtwo{
W=\Tr[A_1E_1B_2C_3-A_2E_1B_1C_3E_2+A_2B_3C_1
-A_3B_2C_1+A_3B_1C_2E_2-A_1B_3C_2]~.
}
Contrary to the case of $dP_1$, 
if we set only the fields $C_i$'s to zero
it is not easy to 
satisfy the F-term condition coming from the 
superpotential \WdPtwo\ with
simultaneously satisfying the D-term condition.
Therefore,
we set $B_i=C_i=0$ to find a classical solution
for fuzzy $dP_2$, which
corresponds to the diagram in Fig. 5b.
Then the F-term condition is automatically satisfied.
The D-term condition for 
the remaining non-zero fields $A_{1,2,3}, E_{1,2}$
reads
\eqn\DtermdPtwo{\eqalign{
\zeta_4{\bf 1}_{V_4}&=A_1A_1^\dag+E_2E_2^\dag\cr
\zeta_0{\bf 1}_{V_0}&=A_2A_2^\dag+A_3A_3^\dag-E_2^\dag E_2\cr
\zeta_1{\bf 1}_{V_1}&=E_1E_1^\dag-
A_1^\dag A_1-A_2^\dag A_2\cr
\zeta_2{\bf 1}_{V_2}&=-E_1^\dag E_1-A_3^\dag A_3~.
}}
This set of equations is solved by
\eqn\dPtwosol{
A_i=ca_i,\quad E_1=ca_4,\quad E_2=ca_5,
}
where $a_i~(i=1,...,5)$ are five independent oscillators
and the coefficient $c$ is a complex number.
We choose the Chan-Paton spaces as
\eqn\dPtwoVi{
V_4={\cal F}_{N,M_1,M_2},\quad
V_0={\cal F}_{N,M_1,M_2+1},\quad
V_1={\cal F}_{N+1,M_1,M_2+1},\quad
V_2={\cal F}_{N+1,M_1+1,M_2+1},
}
where ${\cal F}_{N,M_1,M_2}$ is defined as
\eqn\FNdPtwo{
{\cal F}_{N,M_1,M_2}=\Big\{|m_1,m_2,m_3,m_4,m_5\ket,\quad
\sum_{i=1}^3m_i=N,\quad
m_3+m_4=M_1,\quad
m_1+m_5=M_2\Big\}.
}
The dimension of the space ${\cal F}_{N,M_1,M_2}$
can be calculated from Fig. 6 and is given by
\eqn\dimFdPtwo{
{\rm dim}\,{\cal F}_{N,M_1,M_2}
=\hf(N+1)(N+2) - \hf(N-M_1)(N-M_1+1) - \hf (N-M_2)(N-M_2+1)~.
}
\fig{``Quantized'' toric diagram of fuzzy $dP_2$.
}{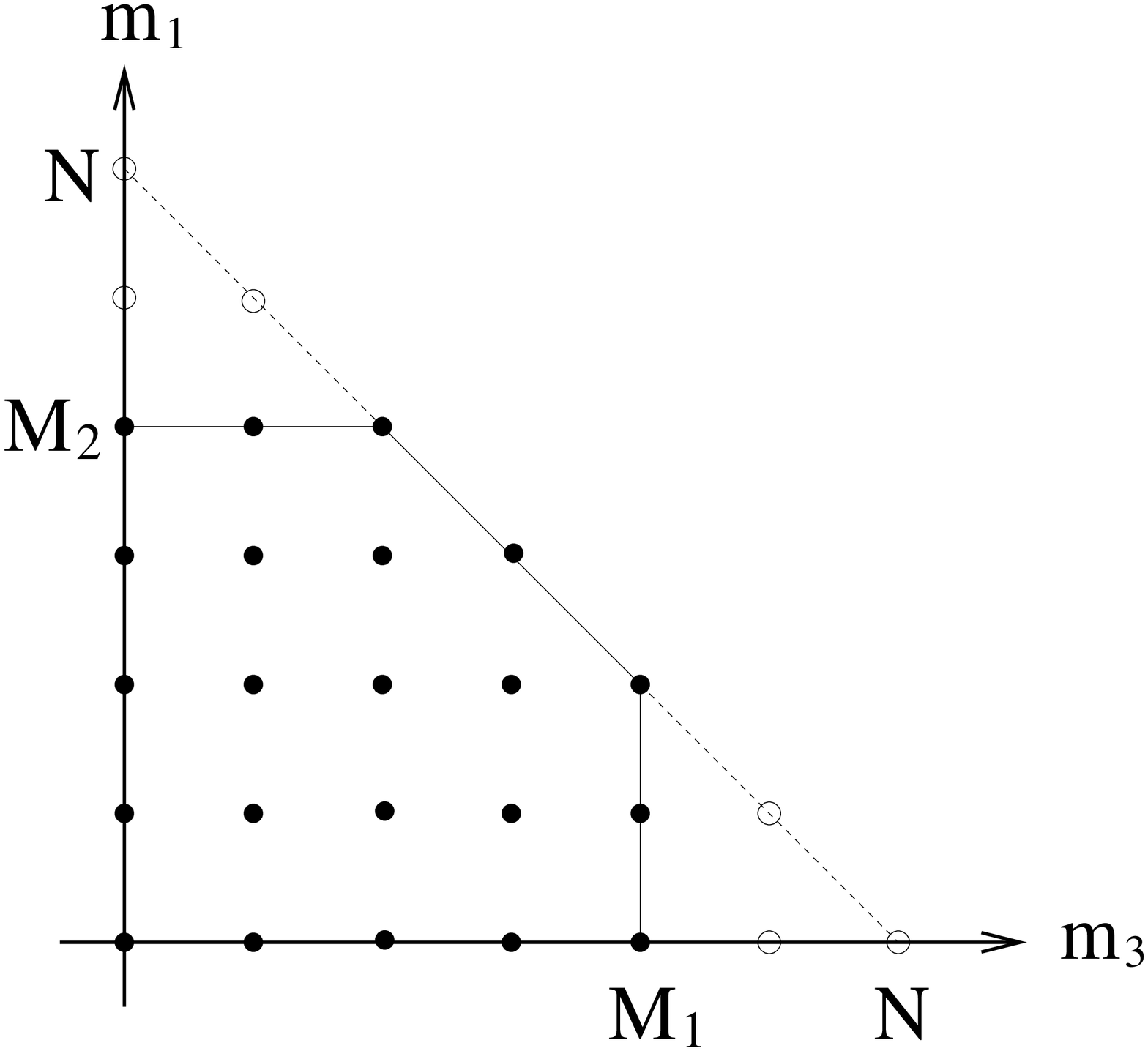}{5cm}
\noindent The FI parameters are quantized as follows
\eqn\FIdPtwo{\eqalign{
\zeta_4&=|c|^2(M_2+2)\cr
\zeta_0&=|c|^2(N-M_2+1)\cr
\zeta_1&=-|c|^2(N-M_1)\cr
\zeta_2&=-|c|^2(M_1+1)~.
}}
Using \dimFdPtwo\ one can show that
\eqn\Finisum{
\sum_{r=0,1,2,4} \zeta_r{\rm dim}\,V_r=0~,
}
which means that the overall $U(1)$ of the gauge group is decoupled.
Let us represent the $U(1)$ subgroup at
the node $r$ by $e^{i \theta_r}$ $(r=0,1,2,4)$.
Since the overall $U(1)$ phase decouples,
we have three $U(1)$ phases left.
We have gauge equivalent relations
\eqn\phaseDPtwo{\eqalign{
A_1 &\sim e^{i (\theta_1-\theta_4)} A_1
= e^{i((\theta_1-\theta_0)+(\theta_0-\theta_4))} A_1 \cr 
A_2 &\sim e^{i (\theta_1-\theta_0)} A_2 \cr
A_3 & \sim e^{i(\theta_2-\theta_0)} A_3
= e^{i((\theta_2-\theta_1)+(\theta_1-\theta_0))} A_3 \cr
E_1 &\sim  e^{i(\theta_2-\theta_1)} E_1 \cr
E_2 &\sim  e^{i(\theta_0-\theta_4)} E_2~,
}
}
thus we can choose $\theta_1-\theta_0$,
$\theta_2 - \theta_1$, $\theta_0-\theta_4$
as three independent $U(1)$ phases.
We observe that \phaseDPtwo\ corresponds
to eq.(37) of \IqbalYE,
while \FNdPtwo\ corresponds
to eq.(36) of \IqbalYE.
Thus our classical solution \dPtwosol\
can be regarded as a fuzzy version 
of the $dP_2$ surface.
As in the case of fuzzy $dP_1$, 
the correspondence of
the space ${\cal F}_{N,M_1,M_2}$
and the toric diagram
of $dP_2$ Fig. 6
can be seen by forgetting the 
dependence on $m_4,m_5$:
\eqn\FNdPtoric{
{\cal F}_{N,M_1,M_2}(dP_2)=
\Big\{|m_1,m_2,m_3\ket,\quad
\sum_{i=1}^3m_i=N,\quad
m_3\leq M_1,\quad
m_1\leq M_2\Big\} .
}
The upper bounds on $m_3,m_1$ mean that the vertices at $|0,0,N\ket$
and $|N,0,0\ket$ are replaced by
line segments representing
two fuzzy ${\Bbb C}{\Bbb P}^1$'s, which appeared
as exceptional cycles
in the blowing up of ${\Bbb C}{\Bbb P}^2$ at two points.

As depicted in  Fig. 7, there is another quiver gauge theory
for the $dP_2$ singularity related to the quiver in
Fig. 5 by Seiberg duality (or toric duality) 
\refs{\FengMI,\FengXR,\BeasleyZP,\BerensteinFI}.
For this quiver gauge theory,
we can also construct a classical solution describing
a D4-brane wrapped on fuzzy $dP_2$
by deleting some of the arrows in Fig 7a.
As in the case of Fig. 5, the F-term condition
is automatically satisfied for the configuration
in Fig. 7b. The D-term condition reads
\eqn\DconddPtwopahseI{\eqalign{
\zeta_0{\bf 1}_{V_0}&=A_1A_1^\dag+E_2E_2^\dag\cr
\zeta_1{\bf 1}_{V_1}&=A_2A_2^\dag -E_1^\dag E_1-E_2^\dag E_2\cr
\zeta_2{\bf 1}_{V_2}&=A_3A_3^\dag+E_1E_1^\dag\cr
\zeta_3{\bf 1}_{V_3}&=-(A_1^\dag A_1+A_2^\dag A_2+A_3^\dag A_3)~.
}}
The solution to these equations is again given by
five independent oscillators in \dPtwosol,
and we choose the Chan-Paton spaces as
\eqn\VforphaseI{
V_0={\cal F}_{N,M_1+1,M_2},\quad
V_1={\cal F}_{N,M_1+1,M_2+1},\quad
V_2={\cal F}_{N,M_1,M_2+1},\quad
V_3={\cal F}_{N+1,M_1+1,M_2+1}.
}
Then the FI-parameters are quantized as follows
\eqn\FIphaseI{\eqalign{
\zeta_0&=|c|^2(M_2+2)\cr
\zeta_1&=|c|^2(N-M_1-M_2-1)\cr
\zeta_2&=|c|^2(M_1+2)\cr
\zeta_3&=-|c|^2(N+1)~.
}}
Again, one can show that
\eqn\Finisum{
\sum_{r=0}^3\zeta_r{\rm dim}\,V_r=0~.
}
The analysis of the $U(1)$ phases is
similar to the previous examples and we will not repeat it here.

\fig{(a) Another quiver diagram for the worldline theory
of D0-branes probing the $dP_2$ singularity.
This is related to the quiver in Fig. 5 by Seiberg duality.
(b) A classical solution for the 
D-branes wrapped on fuzzy $dP_2$ can be obtained
by deleting some arrows in the diagram (a).
}{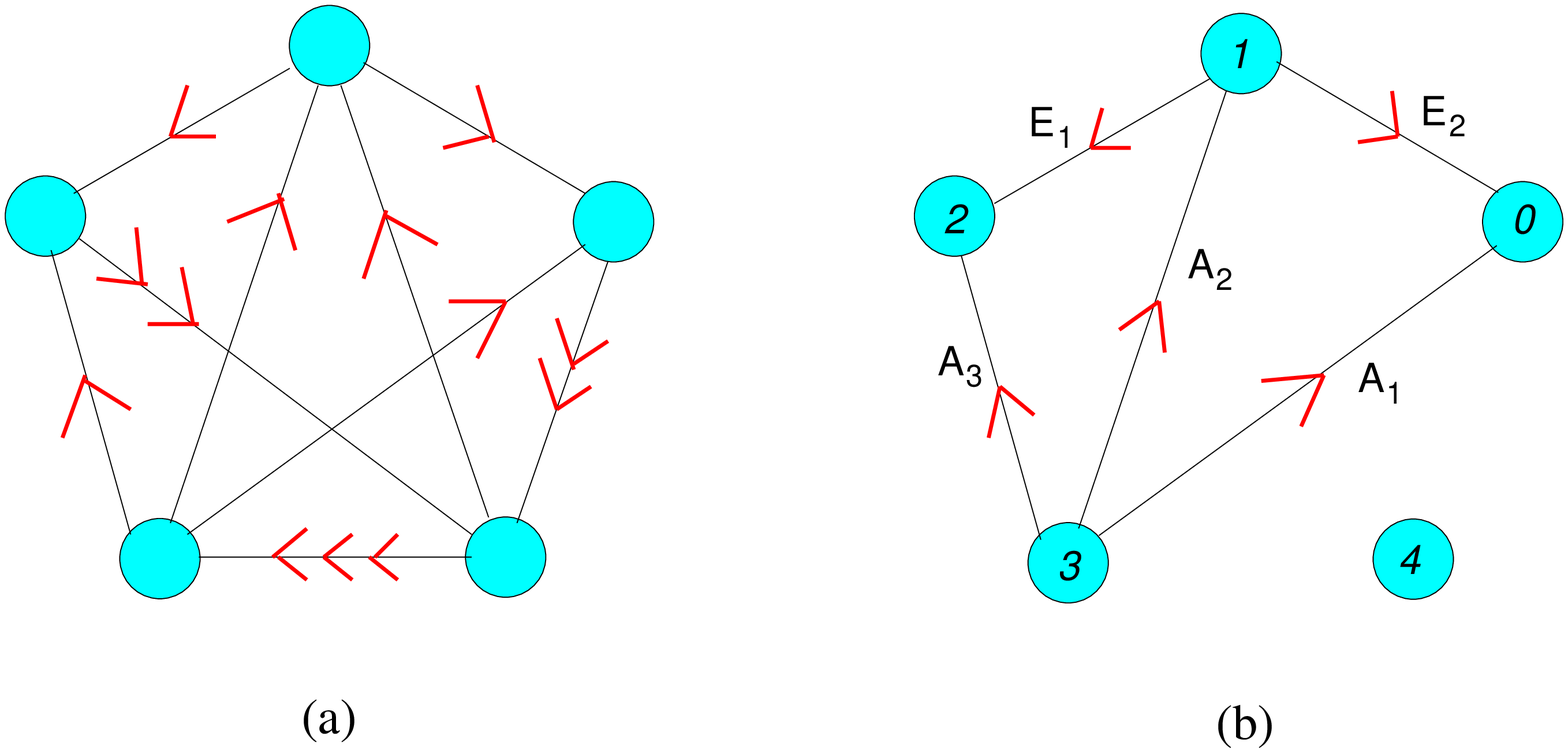}{7cm}

\subsec{Fuzzy dP$_3$}
\fig{(a) The quiver diagram for the worldline theory
of D0-branes probing the $dP_3$ singularity.
(b) A classical solution for the 
D-branes wrapped on fuzzy $dP_3$ can be obtained
by deleting some arrows in the diagram (a).
}{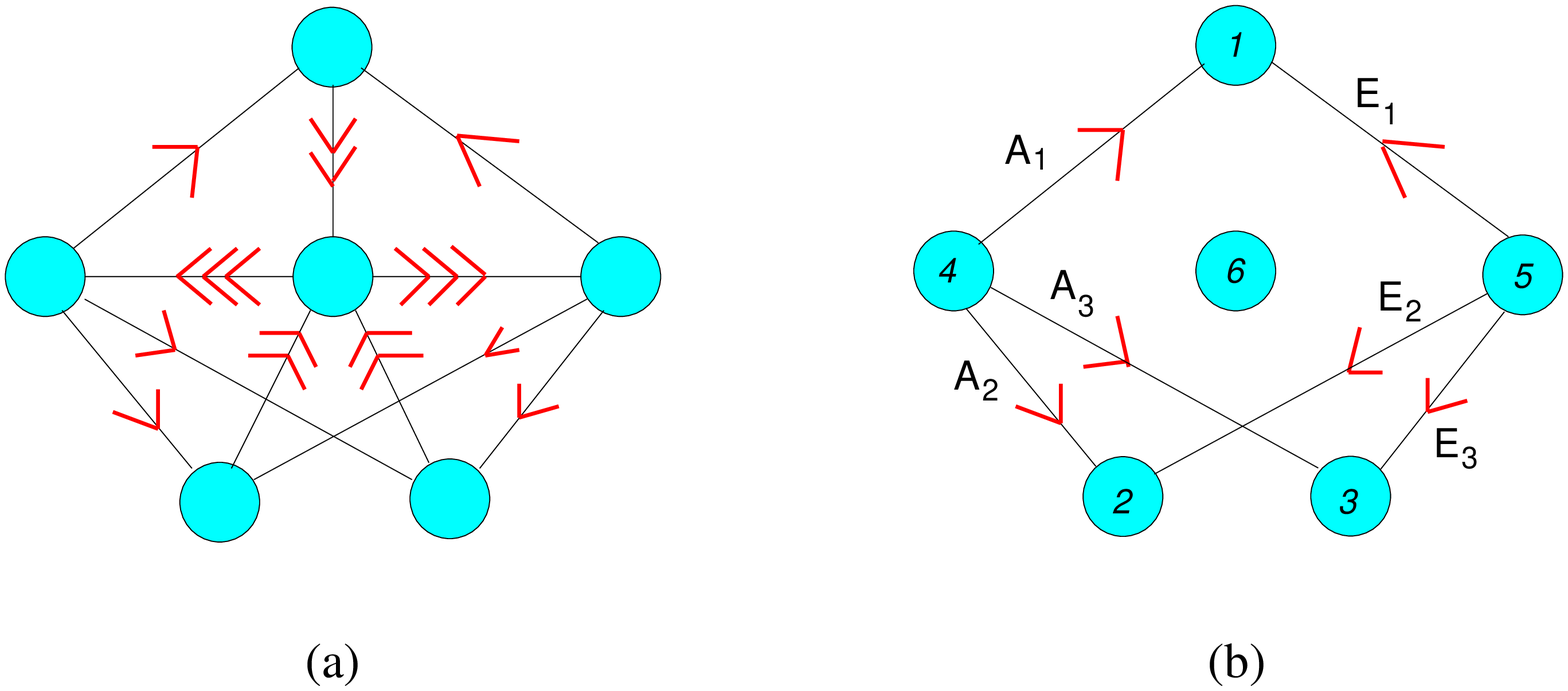}{7cm}

In a manner similar to the previous cases,
we can construct a classical solution describing
a D4-brane wrapped around fuzzy $dP_3$.
Let us consider the quiver diagram for the $dP_3$ singularity
depicted in Fig. 8a, which is called model IV in
\refs{\BeasleyZP,\FengBN,\FrancoRJ}.
A classical solution is
obtained by deleting all arrows connected 
to the node 6 (see Fig. 8b).
We take the remaining non-zero fields as the six independent
oscillators
\eqn\dPthreesol{
A_i=ca_i,\quad E_i=ca_{i+3}\quad(i=1,2,3).
}
For this configuration the F-term condition is satisfied automatically,
and the D-term condition reads
\eqn\dPthreeDcond{\eqalign{
\zeta_i{\bf 1}_{V_i}&=A_iA_i^\dag +E_iE_i^\dag\qquad(i=1,2,3)\cr
\zeta_4{\bf 1}_{V_4}&=-(A_1^\dag A_1+A_2^\dag A_2+A_3^\dag A_3)\cr
\zeta_5{\bf 1}_{V_5}&=-(E_1^\dag E_1+E_2^\dag E_2+E_3^\dag E_3)~.
}}
This is satisfied by taking the Chan-Paton spaces as
\eqn\dPthreeVi{\eqalign{
V_1&={\cal F}_{N,M_1,M_2+1,M_3+1},\quad
V_2={\cal F}_{N,M_1+1,M_2,M_3+1},\quad 
V_3={\cal F}_{N,M_1+1,M_2+1,M_3},\cr
V_4&={\cal F}_{N+1,M_1+1,M_2+1,M_3+1},\quad
V_5={\cal F}_{N,M_1+1,M_2+1,M_3+1}~,
}} 
where 
\eqn\FockdPthree{
{\cal F}_{N,M_1,M_2,M_3}=\lf\{
\prod_{k=1}^6{(a_k^\dag)^{m_k}\o\rt{m_k!}}|0\ket,\quad
\sum_{i=1}^3m_i=N,\quad
m_i+m_{i+3}=M_i~~(i=1,2,3)\ri\}.
}
The dimension of the space ${\cal F}_{N,M_1,M_2,M_3}$
can be calculated from Fig. 9 and is given by
\eqn\dimFdPthree{\eqalign{
{\rm dim}\,{\cal F}_{N,M_1,M_2}
=&\hf(N+1)(N+2) - \hf(N-M_1)(N-M_1+1) \cr
 &- \hf(N-M_2)(N-M_2+1)- \hf(N-M_3)(N-M_3+1)~.
}
}
\fig{``Quantized'' toric diagram of fuzzy $dP_3$.
}{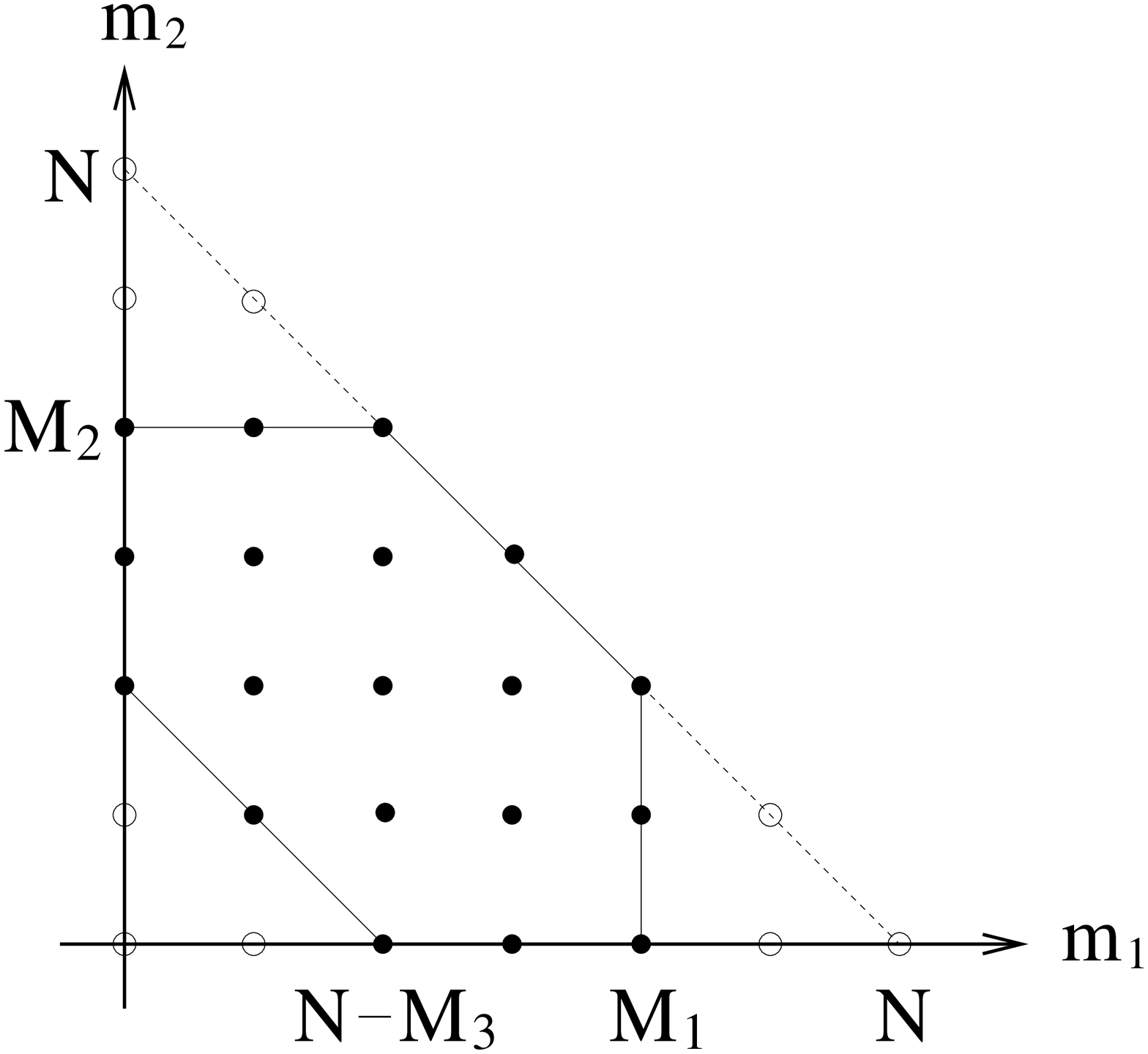}{5cm}
\noindent The FI-parameters are quantized as
\eqn\FIdPthree{\eqalign{
\zeta_r&=|c|^2(M_i+2)\quad(r=1,2,3)\cr
\zeta_4&=-|c|^2(N+1) \cr
\zeta_5&=|c|^2(N-M_1-M_2-M_3-3)~. 
}}
One can show that
\eqn\sumFIdimthree{
\sum_{r=1}^5 \zeta_r {\rm dim}\,V_r=0~,
}
which means that the overall $U(1)$ of the gauge group is decoupled.
Let us represent the $U(1)$ subgroup of
$U(n_r)$ gauge group at
the node $r$ by $e^{i \theta_r}$ $(r=1,2,3,4,5)$.
Since the overall $U(1)$ phase decouples,
we have four $U(1)$ phases left.
We have gauge equivalent relations
\eqn\phaseDPthree{\eqalign{
A_i &\sim e^{i (\theta_4-\theta_i)} A_i 
     = e^{i((\theta_4-\theta_5)+(\theta_5-\theta_i))} A_i \quad (i=1,2,3)\cr 
E_i &\sim  e^{i(\theta_5-\theta_i)} E_i~.
}
}
As before, these equivalence relations through the phases 
\phaseDPthree\ 
together with \FockdPthree\
provides the toric description of $dP_3$.
By forgetting
the dependence of $m_{4,5,6}$, we have
\eqn\FdPthreeeq{
{\cal F}_{N,M_1,M_2,M_3}=\Big\{|m_1,m_2,m_3\ket,\quad
\sum_{i=1}^3m_i=N,\quad
m_i\leq M_i\Big\}.
}
We observe that
the three vertices at $|0,0,N\ket, |0,N,0\ket,|N,0,0\ket$
are replaced by line segments representing 
three fuzzy ${\Bbb C}{\Bbb P}^1$'s,
which appeared as exceptional curves
in the blowing up of ${\Bbb C}{\Bbb P}^2$ at three points.

Some comments on our solution for the $dP_2$ and $dP_3$ cases are in order.
In the case of $dP_0$ and $dP_1$, 
the gauge group is Higgsed completely
around the classical solution if we set the parameters
$c$ and $\til{c}$ both non-zero.
On the other hand, for the $dP_2$ and $dP_3$ cases our
classical solutions have only one parameter $c$, and 
one of the gauge group remains unbroken around the classical
solutions since the corresponding node
is not connected to the
other nodes in 
the quiver diagram for our solution.
Our solutions for $dP_2$ and $dP_3$ 
are probably 
a class of simple solutions 
among more general solutions
for the F-term condition,
similar to $c=0$ (or $\tilde{c}=0$) solutions for
${\Bbb C}{\Bbb P}^2$ and $dP_1$.
Our interpretation
of our solutions for the $dP_2$ and $dP_3$ case is that
they consist of two separated parts. 
One part
represents a D4-brane wrapped around the fuzzy del Pezzo surface,
and the other part associated with the node of the unbroken gauge group
represents 
a collection of D0-branes not bound to
the D4-brane.

Our construction of fuzzy solutions using oscillators,
which is convenient for solving the D-term condition,
works 
if the F-term equations are homogeneous, i.e., they have the
same number of chiral fields on both sides of equation
as in eq.\reducedABeq\ for the ${\Bbb C}{\Bbb P}^2$ case,
and eq.\FforABE\ for the $dP_1$ case.
However, since the superpotentials for the
$dP_2$ and $dP_3$ quiver theories contain
terms of varying degrees, such as cubic, quartic or quintic superpotential
terms etc., many
F-term equations become inhomogeneous 
and hence it is not straightforward to 
solve the F-term condition simultaneously with
the D-term condition by our oscillator method.
Therefore, in the above construction of fuzzy $dP_2$ and $dP_3$
we have deleted many arrows to make 
the F-term equations
trivial (namely, the F-term equations are vacuous $0=0$ for Fig. 7b and Fig. 8b). 
We 
have not yet found a systematic 
method to delete arrows in the $dP_2$ and $dP_3$ quiver
in such a way that the F-term equations are non-vacuous and
at the same time D-term condition is satisfied.
It will be nice to construct
the most general fuzzy $dP_2$ and $dP_3$ solutions 
which will presumably be obtained by
deleting arrows according to the 
Beilinson quiver for $dP_2$ and $dP_3$
(which is sometimes called Bondal quiver \refs{\Bondal,\ProudfootMZ}). 
We will speculate more about the
possible systematic construction of the solutions
in the discussion section.

\newsec{Toward Application to F(uzz) Theory}
So far we have considered classical solutions of 
the guiver gauge theories on D0-branes
describing D4-branes wrapped around fuzzy del Pezzo surfaces.
In the context of local F-theory GUT model building,
it is known that D7-branes should wrap around
a del Pezzo surface in order to be able to take the gravity decoupling limit 
\BeasleyDC. Therefore, it is tempting
to generalize our solutions to D7-branes wrapped on fuzzy del Pezzo surfaces
by starting from D3-branes at del Pezzo singularities instead of
starting from D0-branes. However,
there is a problem of gauge anomaly in the worldvolume theory
of D3-branes if we naively generalize our construction of
fuzzy del Pezzo surfaces to a solution of
the 4-dimensional quiver gauge theory on the fractional D3-branes.
In the next subsection we will comment on possible ways to cancel
the anomaly.
If we put this issue of anomaly aside, 
our construction of D7-branes wrapped on a fuzzy 
del Pezzo surface might have an interesting
applications in the fuzzy version of the F-theory GUTs \HeckmanPV.
 
\subsec{Note on Anomaly Cancellation}
Since the quiver gauge theory in 4-dimensions is a chiral gauge
theory, it has a potential
non-Abelian gauge anomaly.
For the quiver gauge theory
with gauge group $\prod_i U(n_i)$
and the adjacency matrix $a_{ij}$ 
(viz. $a_{ij}$ is the number of chiral multiplets 
in the representation $(n_i,\b{n_j})$),
the condition for the absence of
gauge anomaly at the $i$-th node is\foot{This anomaly cancellation condition 
is closely related to RR-flux conservation 
\CachazoSG.}
\eqn\anomalycond{
\sum_j (a_{ij}-a_{ji})n_j=0~.
}
If we consider the ${\Bbb C}^3/{\Bbb Z}_3$ quiver gauge theory with
our choice of $n_i$ \nithree\ as a 4-dimensional gauge theory,
this anomaly cancellation condition is not satisfied and hence
the gauge symmetry on the fractional D3-brane is anomalous.

At the field theory level,
we can cancel this anomaly by adding extra chiral matter fields 
as ``spectator''. For instance,
in the case of ${\Bbb C}^3/{\Bbb Z}_3$ quiver gauge theory
in Fig. 1a, 
the theory becomes anomaly free if we choose the number
of arrows on each edge to be equal to the rank of gauge group 
at the 
node facing that edge. This cancellation of anomaly
by adding spectator fields and
changing the number of arrows works also for 
the quiver gauge theories associated with higher del Pezzo surfaces. 

In string theory, we cannot arbitrarily add 
chiral matter fields as above.
However, 
we can add flavor D7-branes which wrap a non-compact cycle
in the Calabi-Yau manifold \refs{\FrancoES,\ImamuraFD}
to cancel the anomaly. 
For the case of ${\Bbb C}^3/{\Bbb Z}_3$ quiver gauge theory,
adding flavor D7-branes gives rise to three extra
nodes representing flavor symmetry
groups, and those flavor nodes are connected to
the three nodes of color gauge groups.
One can easily check that for the given gauge
group $U(n_1)\times U(n_2)\times U(n_3)$
the anomaly is canceled
by choosing the rank of the flavor groups appropriately.

The added spectator fields do not play any role in our construction
of classical solution since we can simply set those fields to zero
in the classical solution.
Therefore, we do not have to change the form
of our solution even after adding spectator fields
to make the quiver gauge theory 
anomaly free as a 4-dimensional gauge theory.

\newsec{Discussions}
In this paper, 
we initiated the study of the
classical solutions in quiver gauge theories
which represent D-branes wrapped on fuzzy del Pezzo surfaces.
There will be many interesting directions for further investigations.

For the fuzzy ${\Bbb C}{\Bbb P}^2$ solutions,
we have seen that our construction is closely related 
to the Beilinson's construction of 
stable vector bundles on ${\Bbb C}{\Bbb P}^2$,
and it should be possible
to make similar analysis for the fuzzy $dP_1$ solutions.
On the other hand,
in the construction of the fuzzy $dP_{2,3}$ solutions,
we had quiver diagrams which differ from the Beilinson quivers.
In particular, there was one node which was not included in the diagrams.
This is probably because 
we have picked up a particular simple class of solutions 
for the F-term condition in these cases.
It is known that 
brane tilings \refs{\HananyVE,\FrancoRJ}\foot{%
See \refs{\KennawayTQ,\YamazakiBT} for a review
on brane tilings.}
provides
an efficient way to solve the F-term condition,
and we may be able to find more general solutions
with this technique.
In particular, our construction is based on the toric diagrams,
and therefore it would be applicable
once the F-term condition is rewritten in a form of D-term condition
for a gauged linear sigma model
describing a toric variety, as has been done using brane tilings.
It will also be interesting to investigate
whether our solutions 
with an inclusion of flavor D7-branes to cancel anomaly
have a nice description in the language of brane tilings.

In our construction
the fuzzy del Pezzo surfaces are 
described by finite size matrices,
whose consequences were emphasized in \HeckmanPV.
If we view these toric del Pezzo surfaces
as a fibration of torus over a base space,
the base space is made of finite number of points.
We have seen that
the blow-up at a point in the fuzzy toric del Pezzo surfaces 
amounts to removing some of the points in the toric diagram.
A similar structure has appeared
in the quantum foam description of toric Calabi-Yau manifolds \IqbalDS.
One should note that the physical set up 
there and in this paper are quite different:
The quantum foam describes the fuzziness of the target space-time,
whereas the discussions in this paper was about the 
fuzziness on the worldvolume of D-branes.
Keeping this in mind, 
we may also pursue 
the similarity in the mathematical structures
in both cases.
The mathematical results 
in the study of quantum foam may 
have interesting physical interpretation
in the current setting.
%

While our construction was based on 
the toric description,
it will be nice to have
a generalization of our construction
to include non-toric del Pezzo surfaces.

One of our initial motivations for this work 
was to use D7-branes wrapped on
fuzzy del Pezzo surfaces
in the phenomenological model buildings,
following the idea of \HeckmanPV.
We have presented our preliminary study 
in this direction in section 6.
If we set aside the problem of gauge anomaly, our construction can be
easily generalized to the D3-branes puffed up into D7-branes whose internal
4-dimensional part of the worldvolume 
wraps around a fuzzy del Pezzo surface.
We hope that our construction has useful
applications in the
fuzzy version of F-theory GUT, or F(uzz) theory.
In our approach to F(uzz) theory, 
we start with a 4-dimensional quiver gauge
theory. The fuzzy
extra dimensions arise from the VEV of
the scalar fields 
in the four dimensional theory.
We found that the KK spectrum is similar to that
coming from the usual extra dimensions, but 
there is a UV cut-off for the KK modes set by the non-commutativity.
This opens up a possibility of
studying F(uzz) theory and GUT phenomenology
entirely within a four dimensional 
theory
without introducing extra dimensions from the beginning.
Our scheme can be summarized as
\eqn\fourtoseven{
d=(3+1)~{\rm quiver}~\uerel{{\rm VEV}}{\Longrightarrow}~
{\rm fuzzy}~d=(7+1)~
\uerel{{\rm KK}}{\Longrightarrow}~
d=(3+1)~{\rm GUT}~.
}
In this paper, we introduced the curves as 
data external to the quiver gauge theory, 
but it may be interesting
to realize these curves as
non-commutative solitons \GopakumarZD\
of the theory.
Although the construction in this paper was 
within the framework of type IIB string theory, 
we hope that the essential part 
of our construction also goes through in F-theory.
It would be nice to construct more realistic models
for the F-theory GUT phenomenology using our construction
of fuzzy del Pezzo surfaces.
D-branes wrapped on fuzzy del Pezzo surfaces
may also have applications
in the construction of supersymmetric standard models
from D3-branes on del Pezzo surfaces
\refs{\VerlindeJR,\KrippendorfHJ}.

\vskip7mm
\noindent
\centerline{\bf Acknowledgments}
K.F. is supported 
by the National Center for Theoretical Sciences, Taiwan, R.O.C.
He would like to thank Miranda C. N. Cheng, Wu-Yen Chuang
and Keijiro Takahashi
for the nice lectures on relevant topics at Taiwan String Group.
He also thanks Hirotaka Irie, Hiroshi Isono and Pei-Ming Ho
for useful discussions.
K.O. is supported in part by
JSPS Grant-in-Aid for Young Scientists (B) 19740135.
He would like to thank Michael Douglas and Jonathan Heckman
for discussions during the conference
``String Phenomenology 2010'' at College de France, Paris.
The authors would also like to thank the referee of JHEP
for pointing out a subtle point 
in the calculation of the Yukawa couplings 
which was overlooked in the earlier version of this paper.

\appendix{A}{Central Charge of D4-branes Wrapped on ${\Bbb C}{\Bbb P}^2$}
In this appendix, we review the computation of
BPS central charge of D-branes on ${\Bbb C}^3/{\Bbb Z}_3$
\DouglasQW.
The relation between the ranks of gauge groups $n_{1,2,3}$
for the quiver diagram in Fig. 1a and the D-brane charge
is found by analyzing the monodromy of the 
period integral
at the orbifold point.
In terms of the basis $(1,t,t_d)$ of the solution of the Picard-Fuchs equation
with $t_d\sim \hf t^2+{1\o8}$
near the orbifold point, 
the monodromy
matrix is given by \refs{\DouglasQW,\DiaconescuDT}
\eqn\monodroM{
M=\pmatrix{1&0&0\cr -\hf&-2&-3\cr \hf&1&1}~,
}
which satisfies $M^3=1$.
Applying this matrix to $(1,t,t_d)$, we find
\eqn\Monperiod{
M\pmatrix{1\cr t\cr t_d}=\pmatrix{1\cr-\hf-2t-3t_d\cr\hf+t+t_d},\quad
M^2\pmatrix{1\cr t\cr t_d}=\pmatrix{1\cr-1+t+3t_d\cr\hf-t-2t_d}~.
}
By identifying the periods in the fractional brane basis
\eqn\Piforfrac{
\Pi_1=\hf+t+t_d,\quad
\Pi_2=\hf-t-2t_d,\quad 
\Pi_3=t_d~,
}
the central charge becomes
\eqn\ZinPii{
Z=\sum_{i=1}^3n_i\Pi_i={n_1+n_2\o2}-(n_2-n_1)t+(n_1-2n_2+n_3)t_d~.
}
This gives the relation between $n_i$ and the D-brane charge \QDcharge.
 
This relation also follows from the Beilinson quiver for ${\Bbb C}{\Bbb P}^2$.
The Beilinson quiver is associated with the sequence
\eqn\Beilinsonseq{
{\Bbb C}^{n_1}\tens \Om^2(2)~\uerel{X}{\longrightarrow}~
{\Bbb C}^{n_2}\tens \Om^1(1)~\uerel{Y}{\longrightarrow}~
{\Bbb C}^{n_3}\tens {\cal O}
}
and the central charge $Z$ corresponding to this configuration
is 
\eqn\Zinch{\eqalign{
Z&=n_1{\rm ch}\,\Om^2(2)-n_2{\rm ch}\,\Om^1(1)+n_3{\rm ch}{\cal O}\cr
&=n_1e^{-\om}-n_2(3-e^{\om})+n_3\cr
&=n_1-2n_2+n_3+(n_2-n_1)\om+{n_1+n_2\o2}\om^2~.
}}
Here we have used 
\eqn\Omtwo{
\Om^2=K_{{\Bbb P}^2}={\cal O}(-3),\quad
\Om^2(2)={\cal O}(-3)\tens{\cal O}(2)={\cal O}(-1)~,
}
and the relation
${\rm ch}\,\Om^1(1)=3-e^\om$ used in \Zinch\
comes from the exactness of
the (dual) Euler sequence
\eqn\dualEuler{
0~\longrightarrow~\Om_{{\Bbb P}(V)}(1)~\longrightarrow~
V^*\tens{\cal O}~\longrightarrow~{\cal O}(1)~\longrightarrow~0~,
}
\eqn\chOmone{
{\rm ch}\,\Om^1(1)={\rm ch}\big(V^*\tens{\cal O}\big)
-{\rm ch}\,{\cal O}(1)=3-e^\om~.
}
For the $k$ copies of D4-brane solutions \ABksol,
the D-brane charges are given just by $k$ times the above.
More generally, one can embed D4-brane solutions
with different ranks in a block diagonal form.
In this case, the D-brane charges are given just by
the sum of the D-brane charges in each block.
Sometimes the same gauge group allows
several
different combinations for the rank of 
the D4-brane solutions,
but the total D-brane charges remain the same.

Similarly, we can associate
a sequence for the D2-brane wrapped on
${\Bbb C}{\Bbb P}^1$ in the $A_1$ quiver theory
\eqn\seqCPone{
{\Bbb C}^{n_1}\tens \Om^1(1)~\uerel{X}{\longrightarrow}~
{\Bbb C}^{n_2}\tens {\cal O}~.
} 
The central charge for this configuration is
\eqn\ZforDfive{\eqalign{
Z&=-n_1{\rm ch}\,\Om^1(1)+n_2{\rm ch}\,{\cal O}\cr
&=-n_1{\rm ch}\,{\cal O}(-1)+n_2{\rm ch}\,{\cal O}\cr
&=-n_1e^{-\om}+n_2~.
}}
For our choice of Chan-Paton spaces \nifortwo,
we find
\eqn\ZforAone{
Z=-(N+1)e^{-\om}+N+2=1+(N+1)\om~.
}
This implies that a magnetic flux ${F\o2\pi}=(N+1)\om$
is threading the ${\Bbb C}{\Bbb P}^1$
part of the D2-brane worldvolume, and the quantized sections 
of the line bundle
${\cal O}_{{\Bbb P}^1}(N+1)$ is identified with the Fock space
${\Bbb C}^{n_2}={\cal F}_{N+1}^{(2)}$.
More physically, those states correspond to the lowest Landau levels on
$S^2$ in the presence of magnetic field.

\appendix{B}{Intersections of Curves on Fuzzy ${\Bbb C}{\Bbb P}^2$}

In this appendix, we will consider 
intersections of curves on fuzzy ${\Bbb C}{\Bbb P}^2$.
The curves represent the intersection of other D7-branes
with our fuzzy D7-branes.
In this paper, we treat these curves
as data external to the theory on the fuzzy D7-branes.
We will follow the 
analogous computation for the 
fuzzy ${\Bbb P}^1\times{\Bbb P}^1$ considered in \HeckmanPV.

In terms of the inhomogeneous coordinates
$u=z_1/z_3,~v=z_2/z_3$ of ${\Bbb C}{\Bbb P}^2$
on the patch $z_3\not=0$,
the configuration of matter curves
can be modeled by the Higgsing of
$U(3)$ gauge theory 
on the D7-branes intersecting with the fuzzy D7-branes
down to
$U(1)^3$ according to the background VEV \HeckmanPV
\eqn\phivev{
\Phi_0=\pmatrix{u&0&0\cr 0&0&0\cr 0&0&\beta v}~.
}
Here, we have introduced a complex number $\beta$
which parametrizes the direction of the matter curve.
The three matter curves of this configuration
correspond to
the enhancement of unbroken symmetry $U(1)^3\riya U(1)\times U(2)$
at $u=0, v=0$ and $u= \beta v$.
In terms of the homogeneous coordinates $[z_1:z_2:z_3]$,
the three matter curves are given by
\eqn\matcurve{\eqalign{
{\Bbb C}{\Bbb P}^1_{\rm I}&=\{z_1=0\}\cr
{\Bbb C}{\Bbb P}^1_{\rm II}&=\{z_2=0\}\cr
{\Bbb C}{\Bbb P}^1_{b}&=\{z_1=  \beta z_2\}~.
}}
Yukawa couplings come from
the intersection of the matter curves.
In the commutative case, the intersection of two curves is a single point
for all pair of the curves in \matcurve 
\eqn\intCPone{
{\Bbb C}{\Bbb P}^1_{\rm I}\cap {\Bbb C}{\Bbb P}^1_{\rm II}
={\Bbb C}{\Bbb P}^1_{\rm I}\cap {\Bbb C}{\Bbb P}^1_{b}
={\Bbb C}{\Bbb P}^1_{\rm II}\cap {\Bbb C}{\Bbb P}^1_{b}
=[0:0:1]
}
However, the intersection of those curves is
different from \intCPone\ in the fuzzy setup as we will see below.

On fuzzy ${\Bbb C}{\Bbb P}^2$, these matter curves
correspond to
subspaces of the Fock space ${\cal F}_N^{(3)}$
annihilated by $a_1,a_2$ and $a_1- \beta a_2$, respectively
\eqn\Fockmatcurve{\eqalign{
{\cal F}({\Bbb C}{\Bbb P}^1_{\rm I})&={\rm ker}(a_1)=
\Big\{|0,m_2,m_3\ket,~~m_2+m_3=N\Big\}\cr
{\cal F}({\Bbb C}{\Bbb P}^1_{\rm II})&={\rm ker}(a_2)=
\Big\{|m_1,0,m_3\ket,~~m_1+m_3=N\Big\}\cr
{\cal F}({\Bbb C}{\Bbb P}^1_{b})&={\rm ker}(b_1)=
\Big\{(b_2^\dag a_3)^m|0,0,N\ket,~~0\leq m\leq N\Big\}~,
}}
where we have defined $b_{1},b_2$ as the linear combination
of $a_1$ and $a_2$
\eqn\bonetwo{
b_1={a_1-\beta a_2\o\rt{1+|\beta|^2}},\quad
b_2={\b{\bt}a_1+a_2\o\rt{1+|\beta|^2}},\quad
[b_i,b_j^\dag]=\cob_{ij}~.
}

In the fuzzy setup, the Yukawa coupling is computed by sandwiching
certain operators between projection
operators corresponding to the above
spaces ${\cal F}({\Bbb C}{\Bbb P}^1_{\al})~(\al={\rm I},{\rm II},b)$
\eqn\projpi{\eqalign{
{\cal P}_{\rm I}&=\sum_{m=0}^N|0,m,N-m\ket\bra 0,m,N-m|\cr
{\cal P}_{\rm II}&=\sum_{m=0}^N|m,0,N-m\ket\bra m,0,N-m|\cr
{\cal P}_{b}&=\sum_{m=0}^N|0,m,N-m\ket_b\,{}_b\bra 0,m,N-m|~.
}}
Here $|0,m,N-m\ket_b$ denote the orthogonal 
states in ${\cal F}({\Bbb C}{\Bbb P}^1_{b})$,
i.e., they are the
Fock states of 
the $b_1,b_2,a_3$ oscillators
\eqn\boscibasis{
|m_1,m_2,m_3\ket_b={(b_1^\dag)^{m_1}(b_2^\dag)^{m_2}(a_3^\dag)^{m_3}\o
\rt{m_1!m_2!m_3!}}|0,0,0\ket~.
}
The state $|0,m,N-m\ket_b$ is
written in the original Fock states
of $a_1,a_2,a_3$ oscillators as
\eqn\basispsi{
|0,m,N-m\ket_b=\sum_{k=0}^m {\beta^k \o {\rt{(1+|\beta|^2)^m}}}  \rt{m!\o k!(m-k!)}\,\Big|k,m-k,N-m\Big\ket~.
}
From this expression we can read off
the overlap between 
$|0,m,N-m\ket_b$ and the basis
of ${\cal F}({\Bbb C}{\Bbb P}^1_{\rm I})$,
${\cal F}({\Bbb C}{\Bbb P}^1_{\rm II})$
as
\eqn\psioverlap{\eqalign{
\bra 0,m,N-m|0,m',N'-m'\ket_b
=&
\left({1 \o \rt{1+|\beta|^2}}
\right)^m
 \cob_{m,m'}\cob_{N,N'} \cr
\bra m,0,N-m|0,m',N'-m'\ket_b
=&
\left({\bt \o \rt{1+|\beta|^2}}\right)^m
 \cob_{m,m'}\cob_{N,N'}~.
}}
This means that the states on ${\Bbb C}{\Bbb P}^1_b$
can interpolate
the states on ${\Bbb C}{\Bbb P}^1_{\rm I}$ and ${\Bbb C}{\Bbb P}^1_{\rm II}$.
In particular, the product
of projection operators ${\cal P}_b$ and
${\cal P}_{{\rm I,II}}$
is different from the projection to
the classical intersection point $[0:0:N]$
\intCPone:
\eqn\Projprod{\eqalign{
{\cal P}_{\rm I}{\cal P}_{\rm II}&=|0,0,N\ket\bra 0,0,N|\cr
{\cal P}_{\rm I}{\cal P}_{b}&=\sum_{m=0}^N
\left({1 \o \rt{1+|\beta|^2}}\right)^m 
|0,m,N-m\ket\,\,{}_b\!\bra 0,m,N-m|\cr
{\cal P}_{\rm II}{\cal P}_{b}&=\sum_{m=0}^N
\left({\bt \o \rt{1+|\beta|^2}}\right)^m
|m,0,N-m\ket\,\,{}_b\!\bra 0,m,N-m|~.
}}
It is interesting to observe that
the intersection of the fuzzy curves 
${\Bbb C}{\Bbb P}^1_{b}$
and
${\Bbb C}{\Bbb P}^1_{\rm I,II}$
samples all the ``fuzzy points'' 
on ${\Bbb C}{\Bbb P}^1_{b}$,
i.e. all the states in ${\cal F}({\Bbb C}{\Bbb P}^1_{b})$
\HeckmanPV.
This is different from the situation in the ${\Bbb P}^1\times
{\Bbb P}^1$ example
considered in \HeckmanPV.

If we naively generalize the
computation of Yukawa couplings on ${\Bbb P}^1\times {\Bbb P}^1$
studied in \HeckmanPV\
to our fuzzy ${\Bbb C}{\Bbb P}^2$ case,
we seem to
get non-trivial
Yukawa texture
with a hierarchical factor like $({\beta/\sqrt{1+|\beta|^2}})^m$
coming from the overlap of ${\Bbb C}{\Bbb P}^1_{\rm I,II}$
and ${\Bbb C}{\Bbb P}^1_{b}$ in \psioverlap.
However, 
in the above
the curves representing the 
intersections with other D7-branes
were introduced as external inputs
to the quiver gauge theory,
and it remains to be checked 
whether those curves correctly 
describe the intersections with other D7-branes.
Accordingly, there are some subtle points
in applying 
the naive generalization of the prescription
given in \HeckmanPV\ to our case.\foot{
We would like to thank the referee of JHEP for pointing out some issues. 
}
It might be the case that
our non-holomorphic non-commutativity
does not alter the result of the
holomorphic calculation of the Yukawa couplings in \CecottiZF,
in particular, the rank one theorem in the commutative case.
We would like to clarify this point  in the near future 
and see if 
the above interpolating property
of ${\Bbb C}{\Bbb P}^1_b$ has interesting 
physical
consequences in the texture of the Yukawa couplings.

\listrefs
\bye